\documentclass[twocolumn]{aastex6}

\usepackage{epsf}

\begin{document}

\title{Discovery of diffuse dwarf galaxy candidates around M101 }

\author{P. Bennet\altaffilmark{1}, D. J. Sand\altaffilmark{1,2}, D. Crnojevi\'{c}\altaffilmark{1}, K. Spekkens\altaffilmark{3}, D. Zaritsky\altaffilmark{2}, A. Karunakaran\altaffilmark{3} }

\altaffiltext{1}{Physics \& Astronomy Department, Texas Tech University, Box 41051, Lubbock, TX 79409-1051, USA }
\altaffiltext{2}{Department of Astronomy/Steward Observatory, 933 North Cherry Avenue, Rm. N204, Tucson, AZ 85721-0065, USA}
\altaffiltext{3}{ Department of Physics and Space Science, Royal Military College of Canada, PO Box 17000, Station Forces, Kingston, Ontario, K7K 7B4, Canada}

\begin{abstract}

We have conducted a search of a 9 deg$^{2}$ region of the CFHTLS around the Milky Way analog M101 (D$\sim$7 Mpc), in order to look for previously unknown low surface brightness galaxies. This search has uncovered 38 new low surface brightness dwarf candidates, and confirmed 11 previously reported galaxies, all with central surface brightness $\mu$(g,0)$>$23mag/arcsec$^{2}$, potentially extending the satellite luminosity function for the M101 group by $\sim$1.2 magnitudes. The search was conducted using an algorithm that nearly automates the detection of diffuse dwarf galaxies. The candidates small size and low surface brightness means that the faintest of these objects would likely be missed by traditional visual or computer detection techniques. The dwarf galaxy candidates span a range of $-$7.1 $\geq$ M$_g$ $\geq$ $-$10.2 and half light radii of 118-540 pc at the distance of M101, and they are well fit by simple S\'{e}rsic surface brightness profiles. These properties are consistent with dwarfs in the Local Group, and to match the Local Group luminosity function $\sim$10-20 of these candidates should be satellites of M101. Association with a massive host is supported by the lack of detected star formation and the over density of candidates around M101 compared to the field. The spatial distribution of the dwarf candidates is highly asymmetric, and concentrated to the northeast of M101 and therefore distance measurements will be required to determine if these are genuine members of the M101 group.

\end{abstract}


\section{Introduction} \label{sec:intro}

The faint end of the galaxy luminosity function is a critical proving ground for understanding the astrophysics of the $\Lambda$+Cold Dark Matter ($\Lambda$CDM) model for structure formation \citep[see][for a review]{bullock17}, and significant progress is being made  on both theoretical and observational fronts.  For instance, on the theoretical side, the increase in computational power and the sophisticated treatment of baryonic physics in the latest generation of numerical simulations has greatly improved comparisons with dwarf galaxies in the Local Group \citep[e.g.][]{Brooks13,Sawala16,Wetzel16}.  

Observational progress proceeds along several avenues.  In the Local Group, dwarf galaxy discovery is undergoing another boom from wide-field optical surveys such as ATLAS \citep[e.g.][]{Torrealba16a}, the Panoramic Survey Telescope \& Rapid Response System \citep[Pan-STARRS; e.g.][]{Laevens15}, the Dark Energy Survey \citep[DES; e.g.][]{Bechtol15,Koposov15} and the Dark Energy Camera more generally \citep{Martin15,Kim15a,Kim15b,Drlica16}.  The advent of wide-field imagers on 4-m+ telescopes has allowed the search for faint satellites in resolved stars to extend beyond the Local Group \citep[e.g.][among others]{chi13,sand14,sand15,crnojevic14,denija16,toloba16b,Carlin16, smercina17}.  Meanwhile, wide-field HI surveys have led to the discovery of several populations of Local Volume dwarfs or dwarf candidates \citep[e.g.][]{Cannon11,Sand15b,Janesh17,Leisman17}. Wide-field optical spectroscopy of dwarf candidates around Milky Way analogs has also proven effective in discovering associated satellites \citep{geha17}.

Optical searches for `diffuse' dwarfs well beyond the Local Group have undergone a renaissance, although they have long been a subject of study \citep[e.g.][]{Bothun87,Impey88,Impey96,dalcan97}.  Of particular interest have been so-called `ultra-diffuse galaxies' (UDGs), which are large in size ($r_h$$\gtrsim$1.5 kpc), but have a very low surface brightness and stellar mass, typically $\sim$10$^8$ $M_{\odot}$. This is two orders of magnitude less than normal galaxies of that size \citep[][who provided the informal UDG definition being used here]{vanDokkum15}. It may be that some UDGs are `failed' galaxies which never fully formed their stellar content as they fell into a cluster environment (\citealt{vanDokkum15}; \citealt{yozin15}). Other models suggest that UDGs form via internal processes: they may represent the `high spin' tail of the normal galaxy distribution \citep{Amorisco16}, or may form due to strong galaxy outflows \citep{DiCintio17}. Distinguishing between these formation mechanisms requires an unbiased survey of UDGs in isolated environments; for instance, significant UDGs in the field would cast doubt on formation mechanisms which require strong environmental effects as their sole origin.

Even beyond the current excitement centered around UDGs, there would be great utility in having a diffuse dwarf galaxy sample  -- not just `ultra' diffuse galaxies -- harvested from wide-field optical surveys that is sensitive to both quenched and star forming populations.  For instance, relatively nearby and recent dwarf galaxy discoveries such as Leo~P \citep{Giovanelli13} and Antlia~B \citep{sand15} were easily seen as diffuse sources in public imaging archives prior to discovery. 
Having a survey tool that is sensitive across all LSB objects is important as this yields more information about the wider galaxy and group luminosity functions than a more selective approach. Such a tool is also more sensitive to potential new classes of LSB objects. Recent searches of large imaging datasets has also yielded significant samples of extremely metal poor galaxies \citep{James15} via tuned searches for blue diffuse dwarf galaxies.

In this paper we present a diffuse dwarf galaxy detection algorithm which we test on a $\sim$9 deg$^2$ region around M101, using data from the Canada-France-Hawaii-Telescope Legacy Survey (CFHTLS).  The ultimate goal is to then apply this algorithm on much larger public imaging datasets to ascertain the properties of diffuse galaxies in the field in a reasonable amount of time. There have been many previous attempts to automate the detection of LSB galaxies, through algorithms or observational techniques, \citep[e.g.][]{blanton11,ferrarese12,vollmer12,merritt14,duc15,vanderBurg16,fliri16,trujillo16,greco17}, and we present a detailed comparison with other recent efforts in  \hyperref[subsec:sim]{$\S$3.2}.

M101, which we will assume is at $D$=7 Mpc \citep{tikhonov15} throughout this work, is an ideal test field for our diffuse dwarf detection algorithm. M101 has a similar mass and disk scale length as the MW \citep{tikhonov15}, although its stellar halo appears to be significantly less massive \citep{vanDokkum14}, making a comparison of its satellite galaxy properties with those seen in the Local Group particularly interesting.  Equally pertinent for the current work, M101 has been the subject of several other recent diffuse dwarf searches \citep[][the first two of these will be referred to as M14 and K15 for the remainder of this work]{merritt14,karachentsev15,Java16,muller17}. These searches have also pointed to the nearby, projected presence of the NGC 5485 group ($D$$\sim$27 Mpc, \citealt{tully16}) which is its own source of diffuse dwarf galaxies \citep[see discussions in][hereafter referred to as M16 and D17 respectively]{merritt16,danieli17}.

This paper describes the discovery of 38 diffuse dwarf candidates in the M101 group by a new detection algorithm, specifically designed for modern wide field imaging surveys. In \hyperref[sec:data]{$\S$2} we describe the data used in this paper. In \hyperref[sec:analysis]{$\S$3} we discuss the creation and testing of this semi-automated algorithm to detect diffuse galaxies. In \hyperref[sec:results]{$\S$4} we present the properties and distribution of the 38 new dwarf candidates, and how they compare to other members of the M101 group. In \hyperref[sec:conclusion]{$\S$5} we discuss these results and our conclusions.

\section{Data Overview} \label{sec:data}

We used data from the Wide portion of the Canada-France-Hawaii-Telescope Legacy Survey (CFHTLS).  
In particular we focussed on a $\sim$3$\times$3 deg$^2$ square dataset from the W3 field which is spanned by the nine 1 deg$^2$ pointings W3$-$1$-$1 to W3$-$3+1, using the nomenclature presented in Figure 4 of \citet{gwyn12}. The typical exposure time for the g-band stacks was 2500s, with a pixel scale of 0.186 arcsec per pixel.  
This region was chosen to be roughly centered on M101 and to approximately match the dwarf galaxy search area of M14, as one of our main goals is to evaluate our dwarf detection algorithm with respect to previous work.

The fields were downloaded directly from the Canadian Astronomy Data Centre, as were the Point Spread Function (PSF) for those image stacks, which were used for measuring dwarf structural parameters and generating simulated dwarfs. The construction and calibration of these stacks utilized the MegaPipe data pipeline \citep{Gwyn08}, and is described in detail by \citet{gwyn11}.  The 50\% completeness for point sources in the W3 fields was $g$$\approx$26.0--26.5 and $r$$\approx$25.7--26.2 mag \citep{gwyn11}.

Data from the Galaxy Evolution Explorer (GALEX) \citep{martin05} were also used to check for UV emission from our identified dwarf galaxy candidates, as this can be a strong indicator of recent star formation.  These data were either part of the All-Sky Imaging Survey (AIS) or Medium Imaging Survey (MIS); see \citet{Morrissey07} for details.  

\section{Dwarf Detection Algorithm} \label{sec:analysis}

\subsection{Detection Algorithm} \label{subsec:algorithm}  

\begin{figure*}
\begin{center}

\includegraphics[width=8cm]{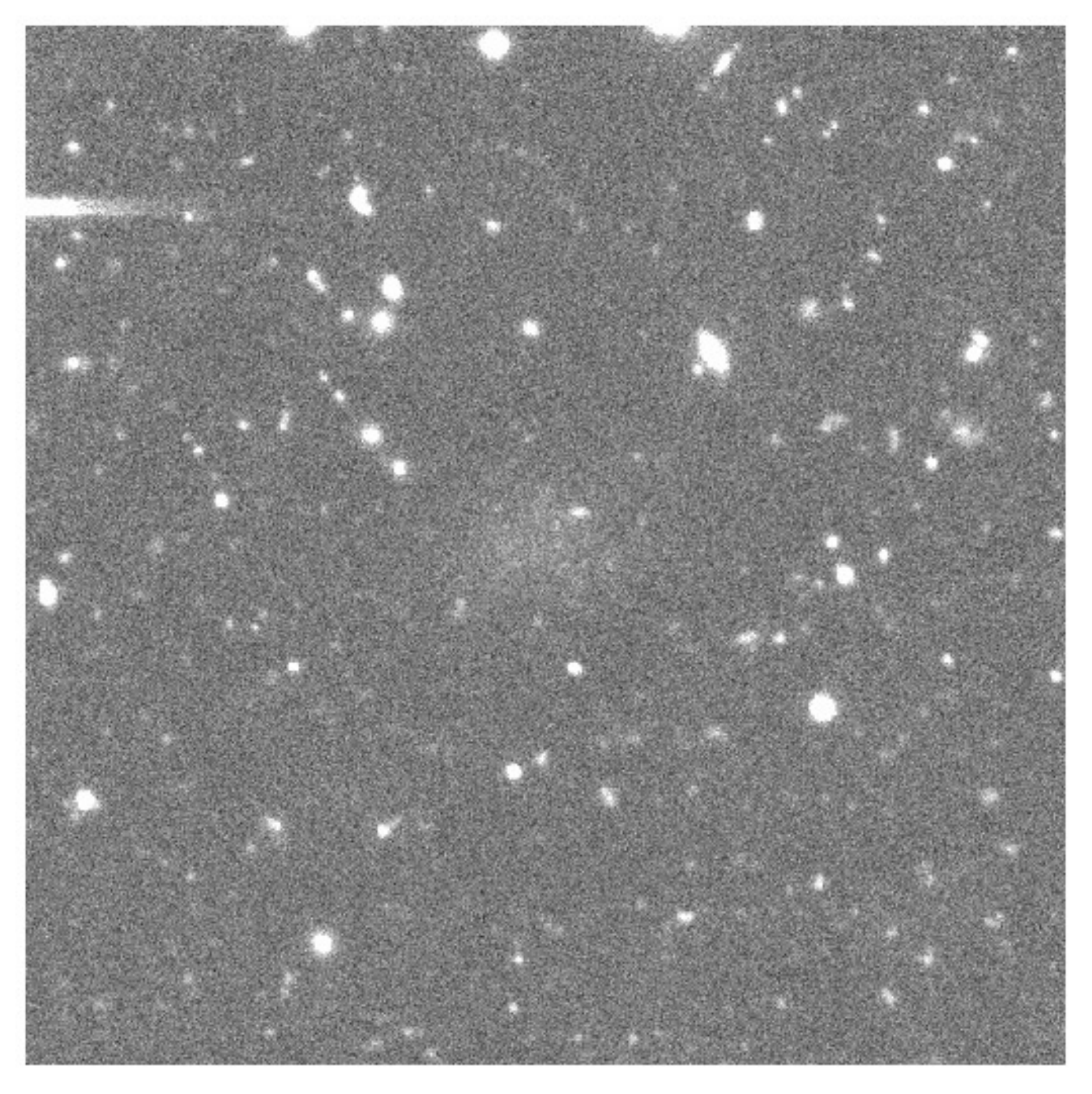}
\includegraphics[width=8cm]{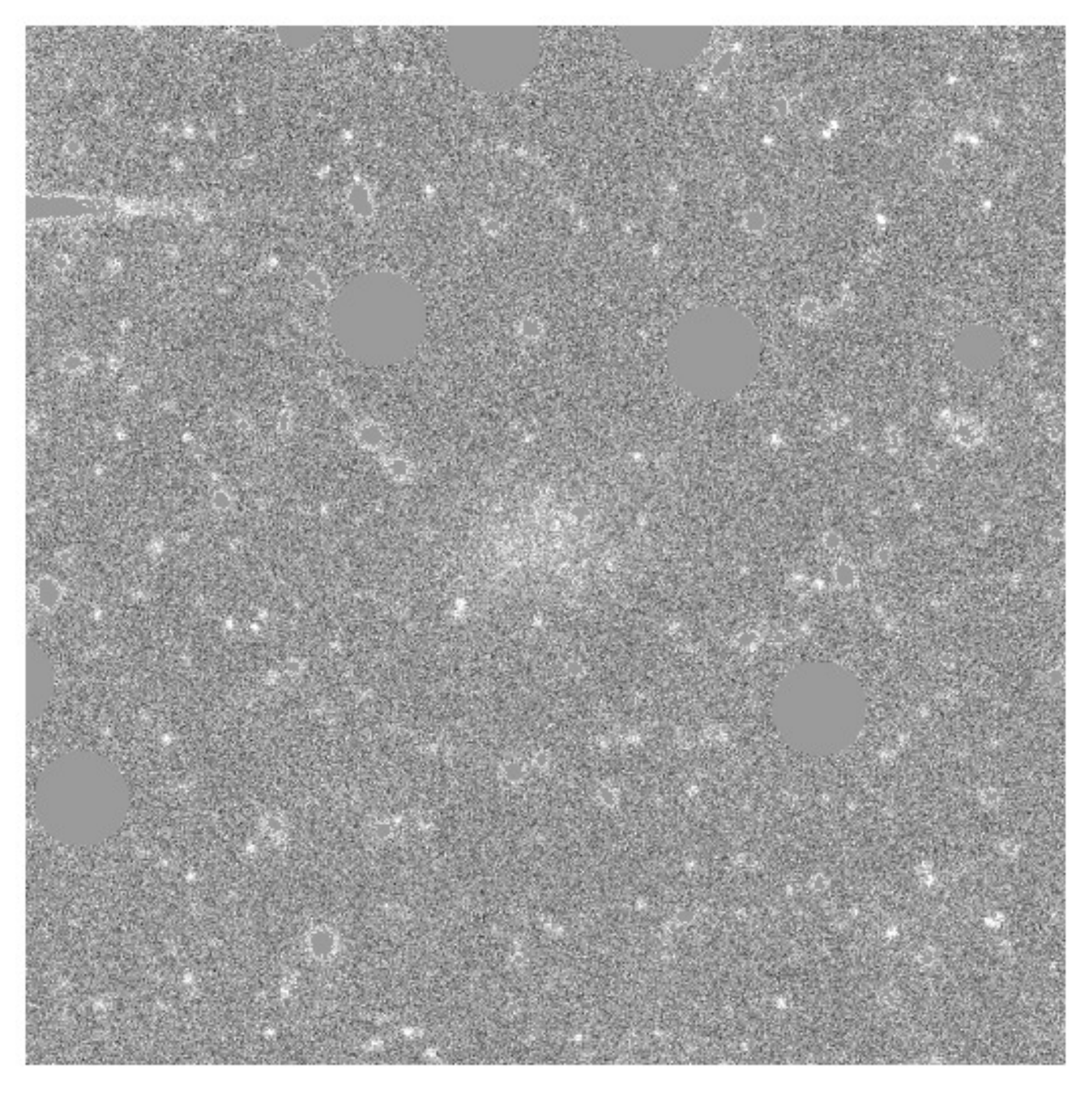}
\includegraphics[width=8cm]{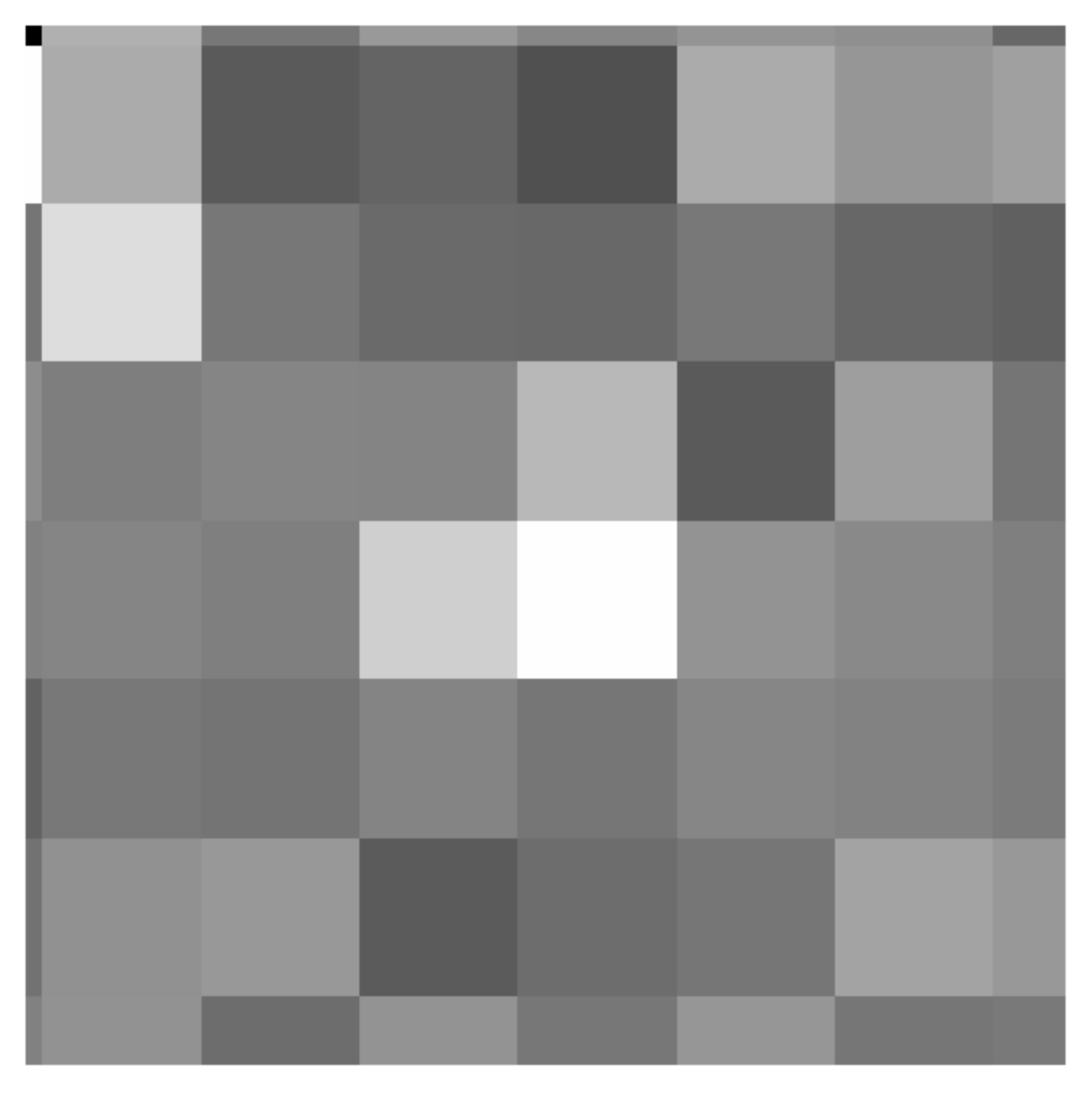}
\includegraphics[width=8cm]{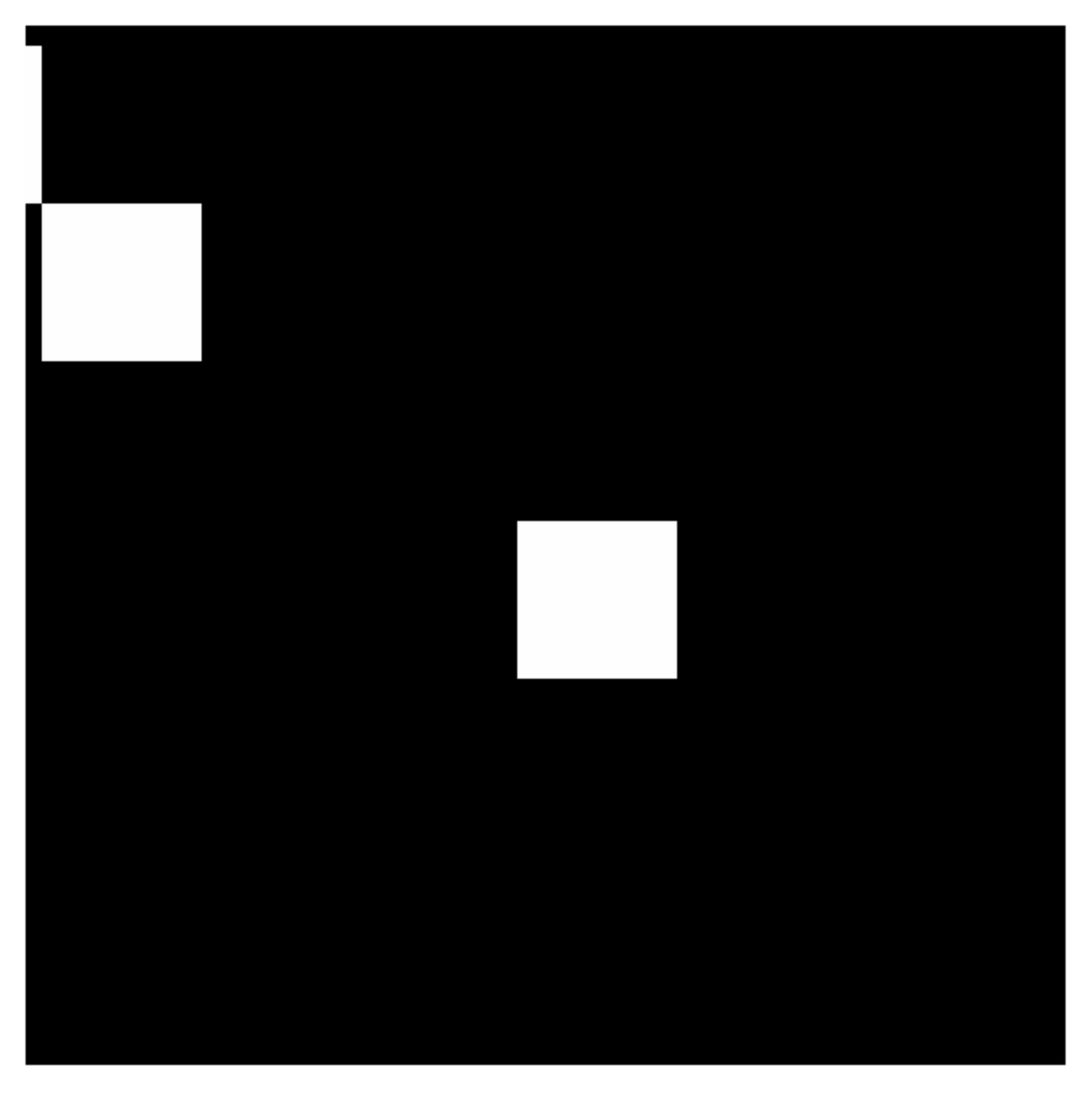}

\caption{A demonstration of the detection algorithm being applied to Dw 9 in the $g$ band of the CFHTLS. North is up, East is left. Panels are 1'$\times$1'. Top Left: Original image of Dw9. Top Right: Image after masking objects from the GSC and a SExtractor catalog of all objects that have  $>$20 pixels at $>$3$\sigma$ above the background. Bottom Left: Image after spatially binning the masked image on a 100x100 pixel scale. Bottom Right: Final extracted objects at  $>$4$\sigma$ in the masked and binned data. Detected objects at this stage are visually inspected to identify final candidates. In addition to Dw 9 a pixel is highlighted to the north east: this is caused by a diffraction spike, a common contaminant easily removed via visual inspection. \label{fig:algor_1}}
\end{center}
\end{figure*}

The new algorithm was developed to detect diffuse dwarf galaxies in g-band images of the CFHTLS and designed to be later applied to other surveys and archival data sets with minimal changes.  A simple illustrative example of the steps involved is shown in Figure~\ref{fig:algor_1}.

The first step in the algorithm utilizes the Guide Star Catalog (GSC) 2.3.2 \citep{lasker08} to mask foreground stars and bright background galaxies. A circular region around each GSC source is masked; the exact size of this region depends on the reported magnitude and was chosen to completely mask the star, although some reflected images remain. This first stage is similar to the initial masking procedure followed in \cite{vanderBurg16}.

Once these bright objects have been masked, the next step is to run SExtractor \citep{bertin96} on the masked image. This identifies sources that have $>$20 pixels at $>$3$\sigma$ above the sky level, such as background galaxies; then these relatively bright pixels are masked as well.
This masking stage does not attempt to remove the extended halos of the SExtracted objects, making this stage less aggressive than the first. Larger background galaxies can have rings of low surface brightness material still visible around the masked area and compact background galaxies or clusters can escape masking entirely (see Figure~\ref{fig:algor_1}). 

Once both masking stages are completed, the algorithm spatially bins the remaining data on a 100$\times$100 pixel scale (corresponding to $\sim$630$\times$630 pc at the distance of M101). This size scale was chosen to maximize the detection of large diffuse objects while remaining sensitive to smaller objects which might correspond to the main locus of dwarf galaxies at the distance of M101. This bin size can be varied to search for dwarf candidates at different size scales, but we keep our bin size fixed in this work.  Extreme outliers among the binned pixels are excluded to remove artifacts that are enhanced by the binning process e.g. chip edges and artificial satellite trails. This outlier removal is very cautious so as not to remove candidates. This binning process allows diffuse objects to be more clearly identified: while background variations even out over large bins, diffuse objects are enhanced and rendered point-like (see bottom left panel of Figure~\ref{fig:algor_1}). 

SExtractor is then run on this binned image and all pixels that are  $>$4$\sigma$ above background are forwarded for visual inspection. This process is shown in full in Figure~\ref{fig:algor_1}.
   
Visual inspection is done via a web interface which simultaneously displays the image outputs of each of the steps described above, along with smoothed image versions of each stage, allowing for easy identification of true diffuse candidates. In a handful of instances, several co-authors examined ambiguous cases together.

The numbers presented in each stage of the above section were arrived at by extensive trials and testing to maximize the parameter space probed by the algorithm.

Typically the final catalogue for a 1 deg$^2$ CFHTLS image around M101 will contain $\sim$200 objects marked for visual inspection, of which $\sim$5--6 are confirmed as strong diffuse dwarf galaxy candidates.  This corresponds to a rate of 1 dwarf candidate per $\sim$35 forwarded objects, which is comparable to or better than other semi-automated detection algorithms presented in the literature \citep[e.g.][]{vollmer12,merritt14,vanderBurg16}. 
False positives were mostly background galaxies and galaxy clusters ($\sim$50\%) or reflection halos around foreground stars ($\sim$40\%), which are especially prominent with CFHT-Megacam.  Other phenomena, such as diffraction spikes and optical ghosts make up the remainder. Another recently published LSB detection algorithm  \citep{greco17} is comprehensively compared in \hyperref[subsec:sim]{$\S$3.2}.

Future refinements to the algorithm -- such as a variable binning size, more aggressive masking of reflections and background galaxy halos, use of color information as a number of false positive sources are limited to single filters, and the addition of a final cut which eliminates detections which are not well fit by modeling software before visual inspection -- will reduce the number of false positives.  Once these refinements have been completed we will publicly release our code. 

\begin{figure*}
\begin{center}
\includegraphics[width=18cm]{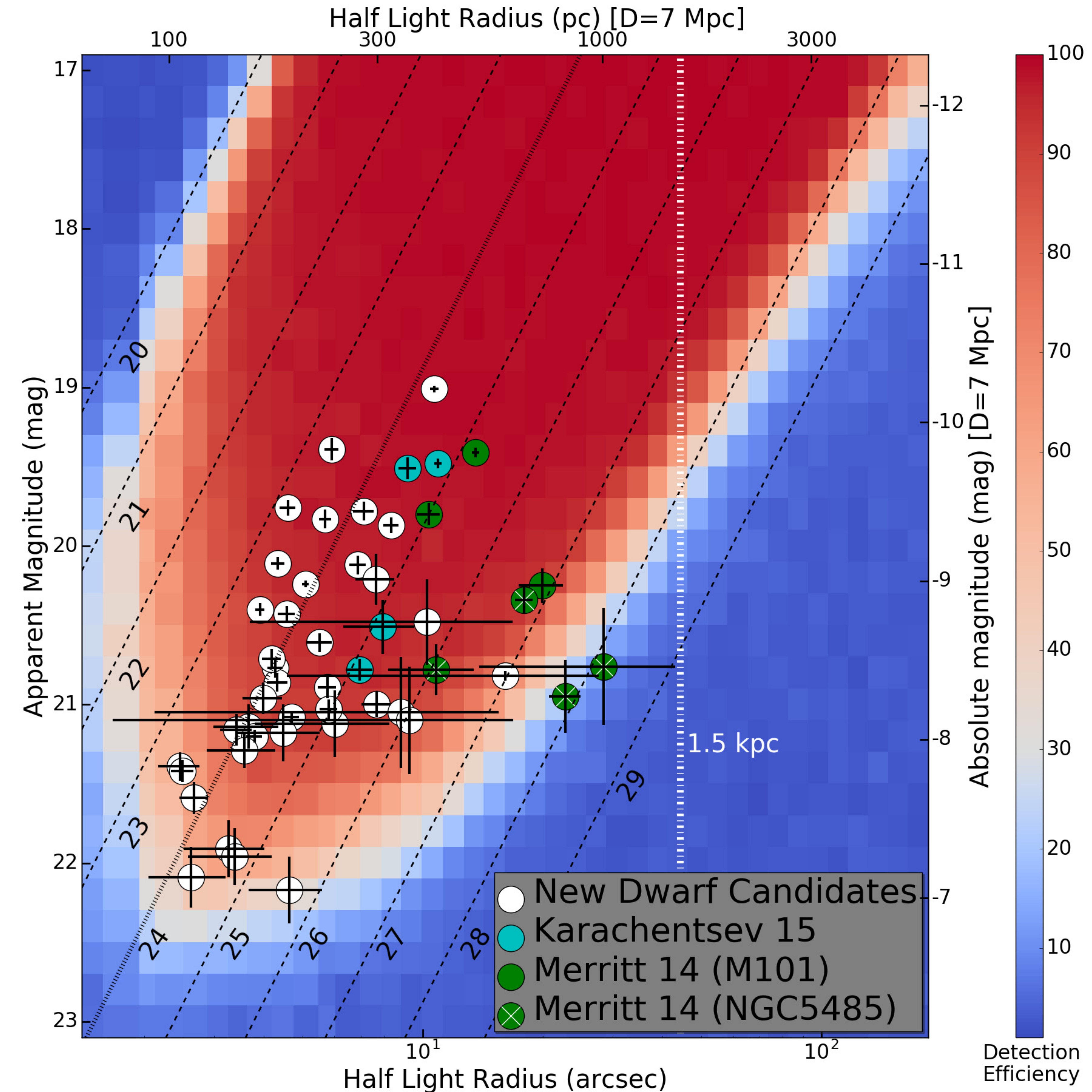}
\caption{Dwarf detection efficiency as a function of magnitude and half light radius.  Redder colors indicate greater detection efficiency (see color scale). 
The white dots indicate the properties of newly discovered dwarfs, cyan dots indicate the dwarfs from K15, green dots indicate the dwarfs from M14 on the apparent properties axes; those with white crosses have been confirmed by M16 to be in the background and do not match the absolute axes. Lines of constant central surface brightness are shown (dashed black).  
The top and right axes assume a distance of 7 Mpc for M101. The dashed white line indicates the boundary for UDGs \citep{vanDokkum15} at the distance of M101, the $\mu$(g,0)$=$24mag/arcsec$^{2}$ line is highlighted as the surface brightness criteria for UDGs, included for completeness even though the UDG phase-space can be defined solely by radius in this diagram. 
\label{fig:det_eff}}
\end{center}
\end{figure*}

\subsection{Simulations} \label{subsec:sim}

To evaluate the effectiveness of our detection algorithm a series of simulated dwarfs were injected into the stacked CFHTLS images around M101. These injected dwarfs were chosen to have a S\'{e}rsic profile \citep{sersic68} with a constant index of  $n$=1, which is typical for diffuse dwarf galaxies \citep{vanderBurg16,vanDokkum15,koda15}, and are injected in a random uniform spatial distribution in batches of 10. 
All simulated dwarfs were placed at least 14.5 arcmin from the center of M101 as any candidates in this region are undetectable due to the projection onto the disk. We did not vary the ellipticity, and only considered circular simulated dwarfs for the main simulation. While moderate ellipticities do not impact the detection efficiency in comparable simulations from other works \citep{vanderBurg16} and a smaller scale test involving $\sim$4$\times$10$^4$ simulated dwarfs showed similar results for ellipticities up to 0.4. We will explore this further in the future.

The injected dwarfs have $g$ band magnitudes between 17 and 23 and half light radii in the range 1.9-165.7 arcsec, which translates to absolute magnitudes between -12.2 and -6.2 and half light radii of 63-5600 pc at the distance of M101. This covers the entire parameter space of the newly detected dwarfs and tests our sensitivity to UDGs at that distance. 
The detection algorithm described in $\S$~\ref{subsec:algorithm} is then used to quantify our detection efficiency, which is detailed in Figure~\ref{fig:det_eff}. 
In total, $\sim$9$\times$10$^5$ simulated dwarfs were injected into our nine CFHTLS search fields. 
As each field has slightly different properties the detection algorithm varies slightly in effectiveness based on the field; Figure~\ref{fig:det_eff} is an average of the 9 fields around M101.

The detection efficiency is $>$90\% for the brighter and more compact dwarfs, with efficiency rapidly falling to below 50\% for dwarfs with $\mu$(g,0)$>$28 mag/arcsec$^{2}$ or m$_g$$>$22. The exception to this are dwarfs which are $\mu$(g,0)$<$20.5 mag/arcsec$^{2}$ for which we have very low probability to detect. This is because they are masked by the algorithm as the central surface brightness exceeds the threshold in the masking stage. We do not consider this to be a large problem as there is a large area between 20.5 and 23 mag/arcsec$^{2}$ of probed parameter space with no detections: if the phase-space distribution of our targets is roughly continuous, then we expect few objects with $\mu$(g,0)$<$23.0 mag/arcsec$^{2}$. This strongly suggests the dwarf population lies in the high detection region with virtually none in the low detection region at brighter magnitudes.

There are however some detected lower surface brightness dwarfs that are outside the high detection region. The M14 objects DF-6 and DF-7 are at the edge of the high detection region, with detection efficiency of $\sim$40\% and $\sim$25\% respectively. This suggests the potential for other similar objects in this field that were undetected.

These detection limits compare favorably with those in \cite{greco17}: these authors present an algorithm led search of LSB objects in $\sim$200 square degrees of the wide layer of the Hyper Suprime-Cam Subaru Strategic Program. The algorithm described in the Greco et. al. paper has a substantially lower false positive rate with roughly half of the final catalogue being LSB candidates; however it also examines a smaller area of magnitude-radius parameter space. Our algorithm has sensitivity to objects with r$_h$$\gtrapprox$100'', whereas the Greco objects all have r$_h$$\leq$14'' and mostly below half that. We also have fainter surface brightness limits with the break between high and low detection regions at $\mu$(g,eff)=29.2 rather than $\mu$(g,eff)=27.4, as is the case in Greco.

\section{Results} \label{sec:results}

We have applied our new diffuse dwarf detection algorithm to the CFHTLS fields around M101, detecting a total of 49 dwarf candidates, 
of which 38 are new and 11 were previously detected in other studies. All dwarf candidates project within the virial radius of M101 (260 kpc, 2.1 deg; M14). Images of our new dwarf candidates can be found in Appendix~\ref{sec:images}. One of our candidates (Dw 26) was detected in follow up HI observations carried out with the Robert C. Byrd Green Bank Telescope, which indicates that this candidate is a massive (M$_{HI}$$\sim$1.21x10$^{9}$  M$_{\odot}$) background galaxy with velocity V$_{sys}$$\sim$11,000 km/s (this leads to a luminosity distance of D$\sim$150 Mpc). The derived velocity confirms that it is an independent background object and not associated with either M101 or NGC 5485 (see Karunakaran et al. in prep. for more details on this and other HI observations of M101 candidates). This leaves 37 dwarf candidates as possible M101 group members.

Of the eleven previously known diffuse dwarf candidates, four are from K15, and seven are from M14.  We remind the reader that four out of the seven M101 dwarf candidates found by M14 are now believed to be members of the background group NGC 5485 at $D$$\sim$27 Mpc (M17, D17), a fact we will return to later on in our discussion. There are no known dwarfs in the M101 group, in our examined magnitude range, that were not detected. 

There are different naming conventions for these objects, with those from M14 referred to as DF1-DF7 and those from K15 as DwA-DwD. We will be using Dw1-38 for our new candidates reported in this paper. 

The detection of 49 dwarf candidates represents a significant over-density with respect to the W3 field as a whole, which yields $\sim$2 strong dwarf candidates per square degree (Bennet et al. in preparation). This over-density strongly suggests that these candidates are associated with either the M101 or NGC 5485 groups.

After cross-checking, we found the majority of our new candidates were too faint to be detected in SDSS, with only 6 of the 38 being detectable. However the detection of the brightest candidates implies that detailed examination of the SDSS could locate LSB objects either in the field or in galaxy groups (e.g., \citealt{trujillo17}). \cite{muller17} indeed performed a visual search of the SDSS in a $\sim330$~deg square region comprising the M101 group. They find several LSB dwarf candidates, but these objects are all brighter than $M_V\sim-10$ and are located beyond the virial radius of M101, thus their search is not directly comparable to ours.
  
To determine if the candidates have ongoing star formation, NUV emission data from the GALEX archive were used. While it is preferable to use FUV to calculate a star formation rate \citep{hao11}, we used NUV because more of the candidates are within the NUV footprint and this allows more consistent characterization across the data set. Almost all candidates within the GALEX footprint show no NUV excess. This lack of NUV emission indicates that these galaxies have an upper limit of $\lesssim$1.7$\pm$0.5x10$^{-3}$  M$_{\odot}$/yr for recent star formation, obtained using the relation from \cite{iglesias06}. From this we can infer that they are passive and are composed of old stellar populations. This is consistent with the MW satellite population, where dwarfs within the virial radius have little gas or ongoing star formation \citep[e.g.][]{spekkens14}. The exception to this is Dw 26, which shows significant NUV excess; we have previously shown that this candidate is a background galaxy.

\subsection{New Dwarf Properties}

We have measured the observational properties for all candidates using GALFIT \citep{peng02}, which can be found in Table \hyperref[tab:obj_prop_1]{1}. We chose a fitting region of 37.2'' (200 pixel) square. 
Any foreground or background sources within this region were masked to minimize contamination. 
The largest objects from M14 required larger fitting regions and for these objects it was also necessary to spatially bin the CFHTLS data in 10x10 pixel bins to ensure that GALFIT could fit them correctly. 

The error bars for the parameters of the dwarf candidates were determined using the procedure from M14. In this procedure, series of 100 dwarfs with parameters identical to each candidate are simulated. These simulated dwarfs are then injected into the image that the candidate was originally found in and their parameters are measured by GALFIT using the same steps as the initial detection. The results obtained from the simulated dwarfs vary due to noise, contamination and systematic errors, and the range of these variations is used to determine the uncertainty in our measurements.

The 37 new dwarf candidates (named Dw 1-38, excluding Dw26) have an apparent g-band magnitude range of 19.0 $\leq$ m$_g$ $\leq$ 22.2 and half light radii between 3-16''. After correction for galactic extinction, this corresponds to an absolute magnitude range of -10.2 $\leq$ M$_g$ $\leq$ -7.1 and half light radii of 118-540 pc at the distance of M101.   
The best fits (see Appendix A) were obtained using a S\'{e}rsic or S\'{e}rsic + Gaussian profile with indices 0.5 $\leq$ $n$ $\leq$ 1.5. S\'{e}rsic profiles have previously been shown to be a good fit for UDGs \citep{vanDokkum15} and dwarf spheroidal galaxies (M14). For those candidates where a simple S\'{e}rsic profile produced extremely high ($>$4) indices, a Gaussian was added to fit possible nuclear structure in several of the dwarf candidates (labelled as nucleated in Table \hyperref[tab:obj_prop_1]{1}). Our search found candidates for the M101 group down to M$_{g}$=-8.2 with 90\% completeness and M$_{g}$=-7.4 with 50\% completeness for galaxies with half light radii $\sim$3''. This is $\sim$1.2 magnitudes fainter than previous surveys of the M101 group (M14, K15). As objects get larger, we become surface brightness, rather than magnitude, limited; the limit moves from $\mu$(g,0)=26 to $\mu$(g,0)=28 depending on candidate radius. See Figure~\ref{fig:det_eff} for a more detailed examination of completeness limits.

The lack of a UDG detection in the M101 group, despite the algorithm being sensitive to this area of parameter space, is in keeping with expectations (\citealt{roman17}, \citealt{vanderBurg17}).
However there are 7 candidates which qualify as UDGs at the distance of the background NGC 5485 group, DF 4-7, Dw A, Dw18 and Dw 32. This would be consistent with the relation between group size and UDG number which predicts $\sim$5 UDGs in the NGC 5485 group.

The newly discovered dwarf candidates fit the trend of the size-luminosity relation from MW and M31 dwarfs reported in \cite{mcconnachie12}, as shown in Figure~\ref{fig:Mag_rad}. Their properties also fit onto the size-luminosity relation if they are shifted to the distance of the background NGC 5485 group, also shown in Figure~\ref{fig:Mag_rad}. That our dwarf candidates fall nicely onto the Local Group size-luminosity relation regardless of the assumed distance implies that assessing group membership solely from this relation can lead to incorrect conclusions.

\subsection{Comparison to Previous Work} 

\begin{figure*}
\begin{center}
\includegraphics[width=8.5cm]{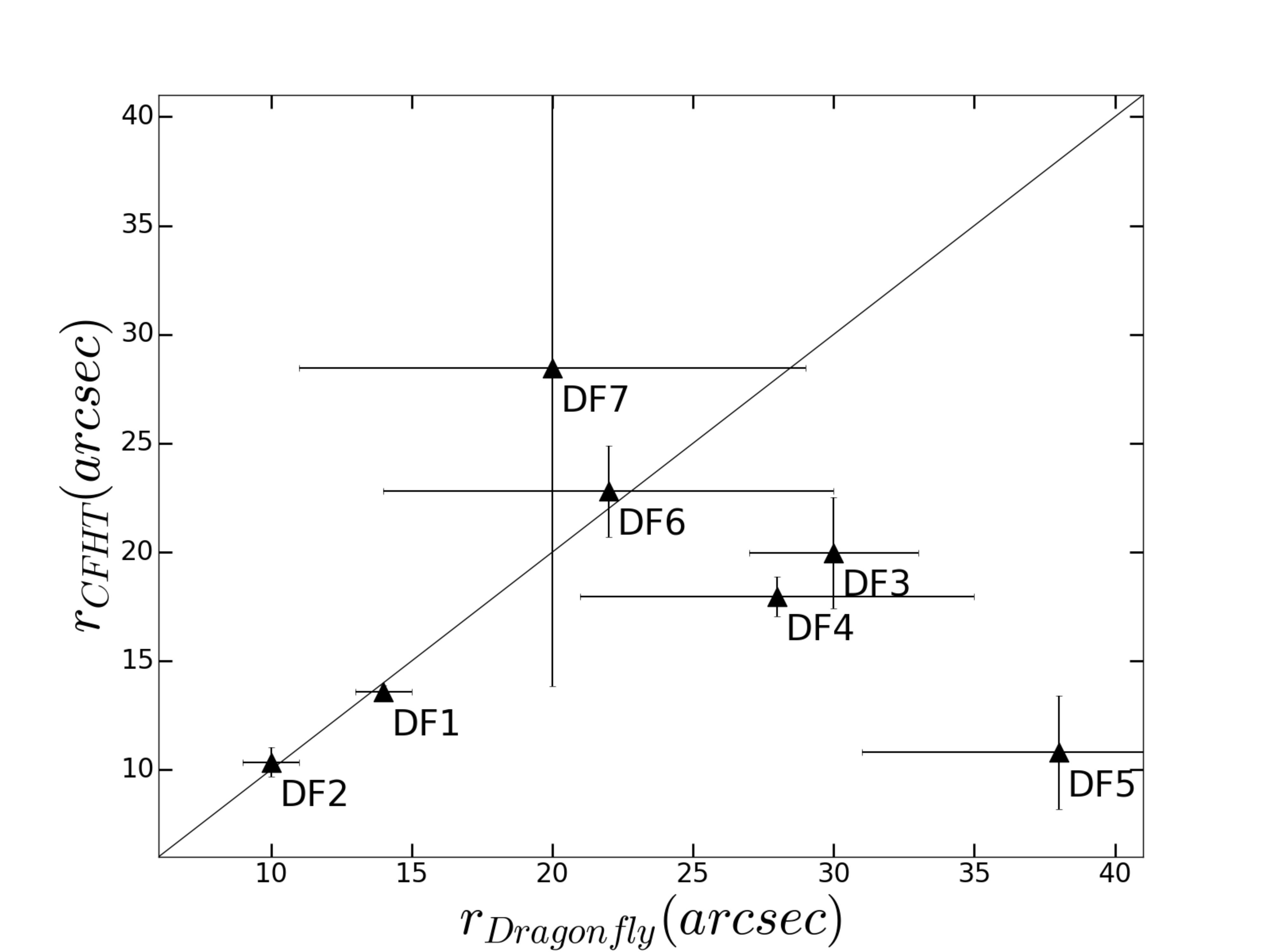}
\includegraphics[width=8.5cm]{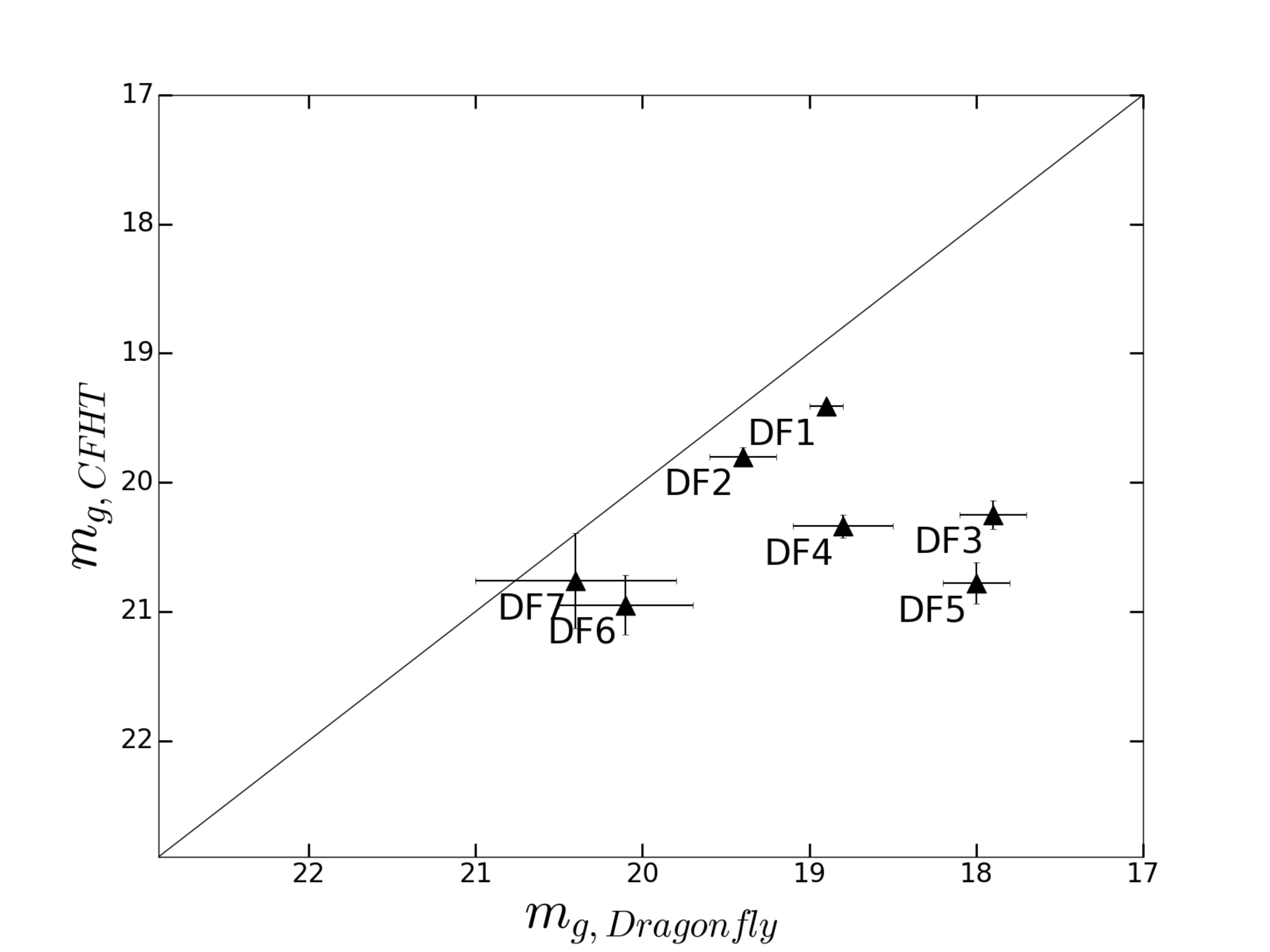}

\caption{A comparison between the reported apparent values for the objects in M14 (Dragonfly) and this work (CFHT). Left: Half light radius. Right: $g$ band magnitude. \label{fig:comp_1}}
\end{center}
\end{figure*}

\begin{figure*}
\begin{center}
\includegraphics[width=18cm]{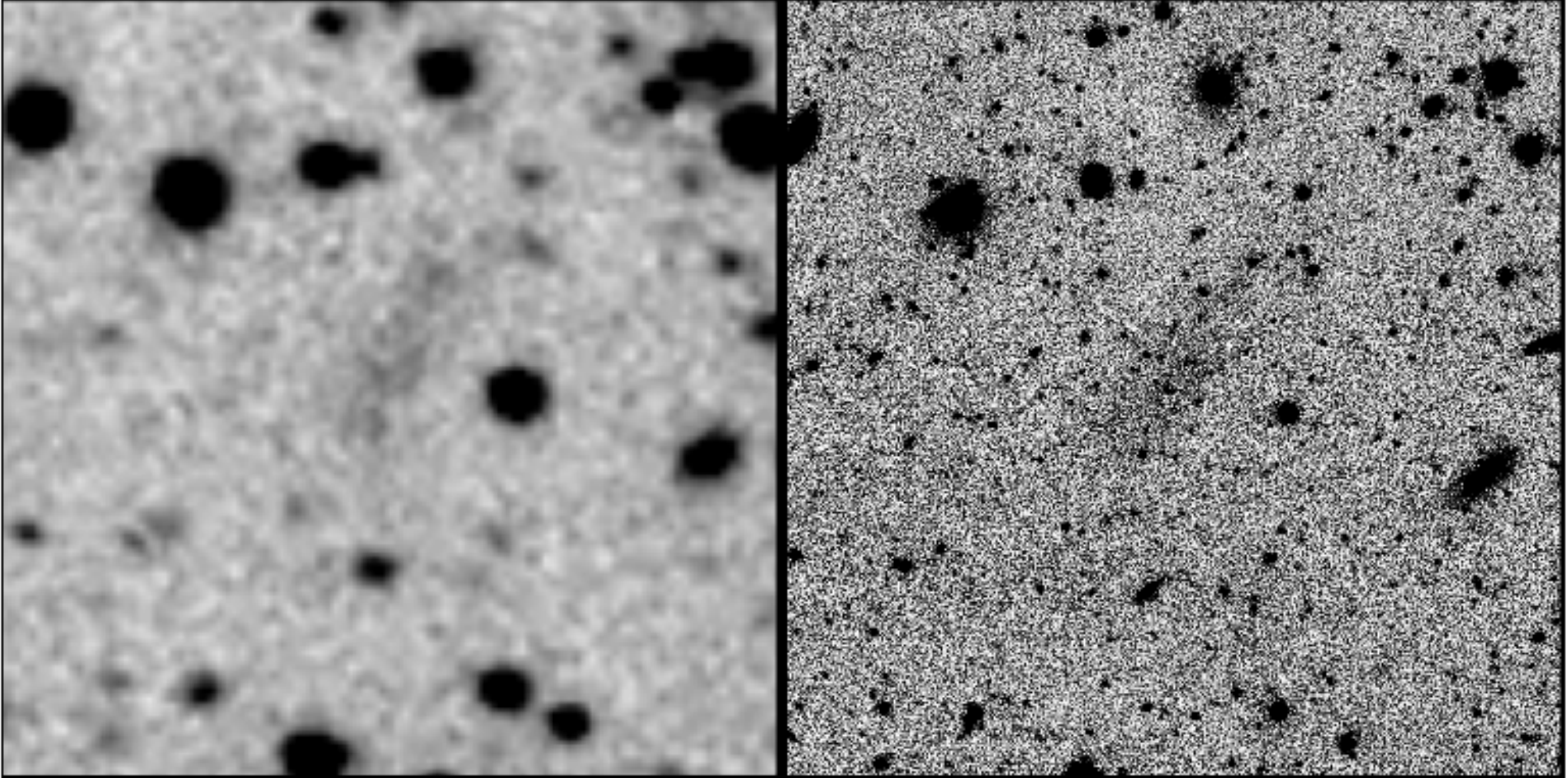}

\caption{This image is from Fig. 6 in M16. A comparison between the r band image for Dragonfly (left) and the CFHTLS (right) of DF-6. Many background objects are clearer and more distinct in the CFHTLS image. Each image is 200''x200''. North is up, east is left.}
\label{fig:comp_2}
\end{center}
\end{figure*}

\begin{figure*}
 \includegraphics[width=0.95\textwidth]{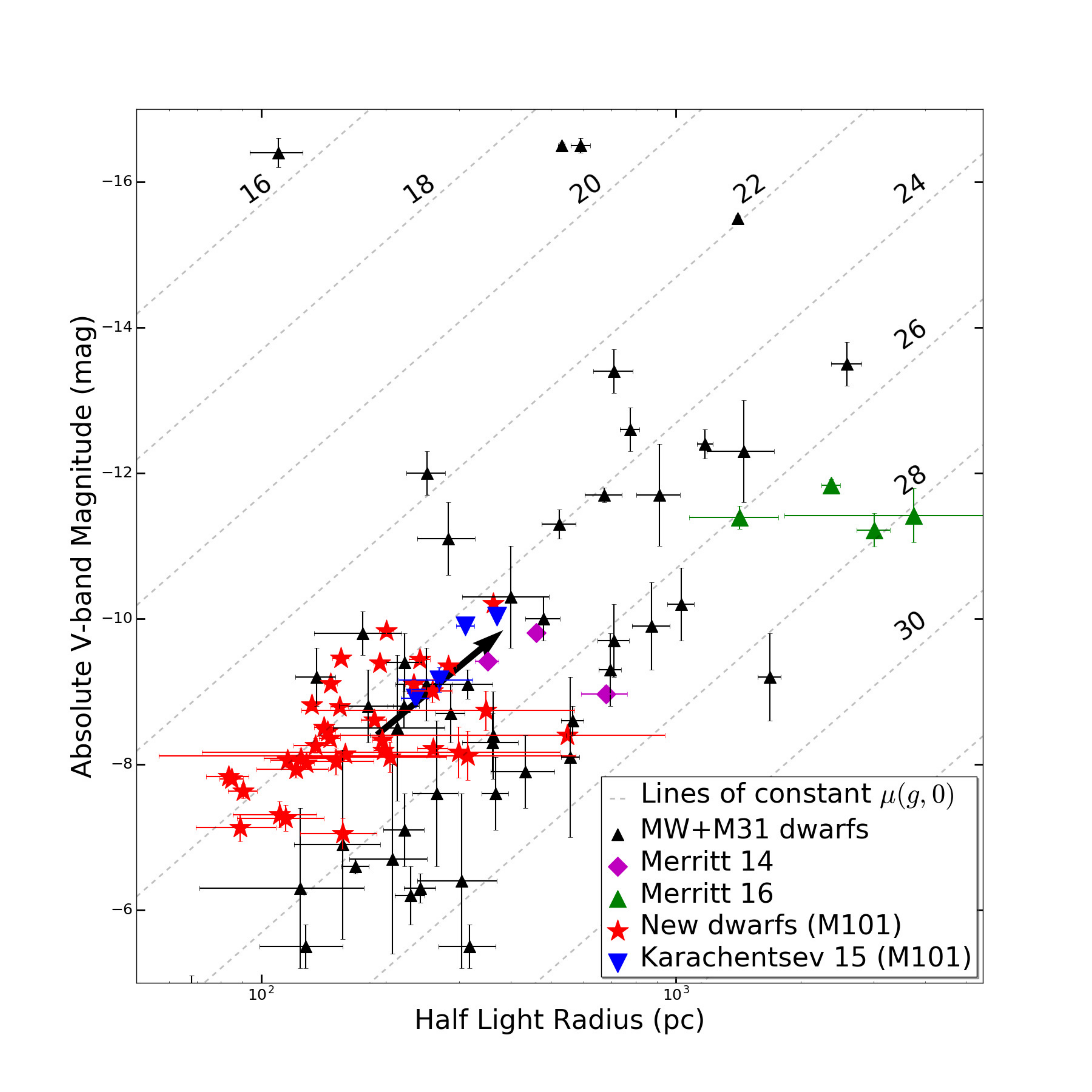}
 \caption{Comparison between the properties of the newly discovered dwarf population around M101 and those of MW and M31 satellites (black triangles; \citealt{mcconnachie12}), as well as those of the M101/NGC 5485 population reported by M14 (magenta diamonds) and M16 (green triangles). Our new dwarf candidates are red stars, assuming a M101 distance (7 Mpc); the black arrow shows the shift in half-light radius and absolute magnitude space if their distance were shifted to that of NGC 5485's (27 Mpc). This shift would also occur for those dwarfs reported in K15 (blue inverted triangles). Lines of constant central surface brightness are shown (dashed grey).\label{fig:Mag_rad}} 
\end{figure*}

Using CFHTLS data, we have obtained the observational properties of the candidates (DF1-7) reported in M14. A comparison between these results and those reported using Dragonfly data is shown in Figure~\ref{fig:comp_1}. The half light radii are largely consistent between the two data sets, with the exception of DF-5, which has HST imaging (M16) that shows a CFHT like radius, though the detection by HST is marginal. This may be due to Dragonfly detecting a LSB halo around DF5 or including unrelated background emission. However the magnitudes derived using the CFHTLS data are consistently fainter than those reported using Dragonfly. A possible explanation is that due to the larger pixel scale of Dragonfly \citep[2.8" vs 0.186" for the CFHTLS;][]{vanDokkum14}, light from background objects is combined with the light from the dwarf galaxy making it appear brighter. This can be seen in Figure~\ref{fig:comp_2}, which is from Fig. 6 in M16 and shown here to illustrate the point on overestimating the luminosity of the dwarfs. To test this hypothesis we degraded the CFHT data to match the optical properties of Dragonfly and reapplied GALFIT: this test resulted in brighter magnitudes than the original CFHTLS data.

\floattable
\begin{deluxetable}{c|cc|c|c|c|c|c|c|c}
\tablenum{1}
\tablecaption{Dwarf Properties \label{tab:obj_prop_1}}
\tablehead{
\colhead{Name} & \colhead{RA} & \colhead{Dec} & \colhead{$g$ band} & \colhead{$r$ band} & \colhead{Half light} & \colhead{Half light} & \colhead{Projected} & \colhead{S\'{e}rsic} & \colhead{Notes}\\
\colhead{} & \colhead{} & \colhead{} & \colhead{Magnitude\tablenotemark{a}} & \colhead{Magnitude\tablenotemark{a}} & \colhead{radius\tablenotemark{a}} &\colhead{radius\tablenotemark{b}} & \colhead{distance from} &  \colhead{Index\tablenotemark{a}} & \colhead{(Confirmed}\\
\colhead{} & \colhead{} & \colhead{} & \colhead{} & \colhead{} & \colhead{(Arcsec)}& \colhead{(pc)} & \colhead{ M101\tablenotemark{b} (kpc)}& \colhead{} & \colhead{associations)}}
\colnumbers
\startdata
DF-1 & 14:03:45.0 & +53:56:40 & 19.4$\pm$0.1 & 18.8$\pm$0.1 & 13.59$\pm$0.29 & 461.2$\pm$9.8 & 50.4 & 0.56$\pm$0.04 & D17, (M101) \\
DF-2 & 14:08:37.5 & +53:19:31 & 19.8$\pm$0.1 & 19.2$\pm$0.1 &10.35$\pm$0.68 & 351$\pm$23 & 96.5 & 0.61$\pm$0.02 & D17, (M101) \\
DF-3 & 14:03:05.7 & +53:36:56 & 20.3$\pm$0.1 & 20.4$\pm$0.1 & 20.0$\pm$2.6 & 678$\pm$87 & 89.6 & 0.50$\pm$0.12 & D17, (M101) \\
DF-4 & 14:07:33.4 & +53:42:36 & 20.3$\pm$0.1 & 20.0$\pm$0.1 & 17.95$\pm$0.92 & 2370$\pm$120\tablenotemark{c} & N/A & 0.23$\pm$0.04 & M16, (NGC 5485) \\
DF-5 & 14:04:28.1 & +53:37:00 & 20.8$\pm$0.2 & 20.6$\pm$0.2 & 10.8$\pm$2.6 & 1420$\pm$340\tablenotemark{c} & N/A & 0.25$\pm$0.15 & M16, (NGC 5485) \\
DF-6 & 14:08:19.0 & +53:11:24 & 21.0$\pm$0.2 & 20.7$\pm$0.2 & 22.8$\pm$2.1 & 3000$\pm$270\tablenotemark{c} & N/A & 0.26$\pm$0.20 & M16, (NGC 5485) \\
DF-7 & 14:05:48.3 & +53:07:58 & 20.8$\pm$0.4 & 21.1$\pm$0.4 & 28$\pm$14 & 3800$\pm$1900\tablenotemark{c} & N/A & 1.40$\pm$0.20 & M16, (NGC 5485) \\
Dw A & 14:06:50.0 & +53:44:29 & 19.5$\pm$0.1 & 19.0$\pm$0.1 & 10.92$\pm$0.23 & 370.6$\pm$7.9 & 98.7 & 0.55$\pm$0.02 & K15 \\
Dw B & 14:08:43.7 & +55:10:02 & 20.8$\pm$0.1 & 20.0$\pm$0.2 & 6.95$\pm$0.54 & 236$\pm$18 & 139.5 & 0.57$\pm$0.08 & K15 \\
Dw C & 14:05:18.2 & +54:53:52 & 20.5$\pm$0.2 & 19.8$\pm$0.2 & 7.9$\pm$1.6 & 269$\pm$55 & 76.6 & 1.05$\pm$0.20 & K15 \\
Dw D & 14:04:24.8 & +53:16:11 & 19.5$\pm$0.1 & 19.2$\pm$0.1 & 9.16$\pm$0.47  & 311$\pm$16 & 133.7 & 0.79$\pm$0.06 & K15 \\
Dw 1 & 14:10:59.7 & +55:53:29 & 20.5$\pm$0.3 & 20.2$\pm$0.2 & 10.3$\pm$6.6 & 350$\pm$220 & 232.5 & 1.56$\pm$0.41 & Nucleated \\
Dw 2 & 14:09:22.0 & +55:18:14 & 20.6$\pm$0.1 & 20.5$\pm$0.1 & 5.52$\pm$0.39 & 187$\pm$13 & 159.3 & 1.23$\pm$0.11 & \\
Dw 3 & 14:08:45.8 & +55:17:14 & 19.8$\pm$0.1 & 19.3$\pm$0.2 & 7.11$\pm$0.42 & 241$\pm$14 & 150.7 & 1.16$\pm$0.07 & \\
Dw 4 & 14:13:01.7 & +55:11:16 & 20.1$\pm$0.1 & 19.9$\pm$0.1 & 6.88$\pm$0.33 & 233$\pm$11 & 201.2 & 0.80$\pm$0.08 & \\
Dw 5 & 14:04:13.0 & +55:43:34 & 20.2$\pm$0.2 & 20.1$\pm$0.1 & 7.64$\pm$0.86 & 259$\pm$29 & 169.2 & 1.33$\pm$0.18 & \\
Dw 6 & 14:02:20.1 & +55:39:17 & 19.9$\pm$0.1 &19.4$\pm$0.1 & 8.34$\pm$0.37 & 283$\pm$12 & 160.3 & 0.83$\pm$0.06 & Nucleated\\
Dw 7 & 14:07:21.0 & +55:03:51 & 21.1$\pm$0.1 & 19.4$\pm$0.1 & 4.70$\pm$0.20 & 159.4$\pm$6.7 & 113.9 & 0.57$\pm$0.06 & \\
Dw 8 & 14:04:24.9 & +55:06:13 & 19.8$\pm$0.1 & 19.3$\pm$0.1 & 5.70$\pm$0.20 & 193.4$\pm$6.8 & 94.6 & 0.62$\pm$0.03 & Nucleated\\
Dw 9 & 13:55:44.6 & +55:08:45 & 21.0$\pm$0.1 & 20.6$\pm$0.1 & 7.66$\pm$0.64 & 260$\pm$22 & 163.7 & 0.52$\pm$0.07 & \\
Dw 10 & 14:01:40.4 & +55:00:57 & 22.2$\pm$0.2 & 21.9$\pm$0.1 & 4.63$\pm$0.97 & 157$\pm$33 & 85.9 & 0.55$\pm$0.19 & \\
Dw 11 & 14:10:04.8 & +54:15:29 & 20.8$\pm$0.1 & 20.1$\pm$0.2 & 4.26$\pm$0.19 & 144.6$\pm$6.3 & 123.0 & 0.50$\pm$0.05 & \\
Dw 12 & 14:09:26.0 & +54:14:51 & 20.9$\pm$0.1 & 20.3$\pm$0.1 & 4.32$\pm$0.25 & 146.6$\pm$8.6 & 111.7 & 1.02$\pm$0.10 & \\
Dw 13 & 14:08:01.2 & +54:22:30 & 20.4$\pm$0.1 & 19.8$\pm$0.1 & 3.91$\pm$0.10 & 132.6$\pm$3.5 & 85.7 & 0.60$\pm$0.03 & \\
Dw 14 & 14:11:03.2 & +53:56:50 & 20.9$\pm$0.1 & 21.6$\pm$0.1 & 5.76$\pm$0.30 & 196$\pm$10 & 148.7 & 0.47$\pm$0.05 & \\
Dw 15 & 14:09:17.5 & +53:45:30 & 21.1$\pm$0.4 & 20.2$\pm$0.1 & 8.8$\pm$6.7 & 300$\pm$230 & 130.7 & 1.98$\pm$0.73 & \\
Dw 16 & 14:03:37.8 & +55:39:51 & 21.2$\pm$0.2 & 20.6$\pm$0.2 & 4.5$\pm$1.0 & 152$\pm$35 & 160.9 & 1.42$\pm$0.25 & \\
Dw 17 & 13:59:12.8 & +55:35:39 & 21.1$\pm$0.1 & 20.6$\pm$0.1 & 3.66$\pm$0.68 & 124$\pm$23 & 167.5 & 1.28$\pm$0.19 & \\
Dw 18 & 13:58:52.3 & +55:29:27 & 20.8$\pm$0.3 & 21.4$\pm$0.2 & 16$\pm$12 & 550$\pm$390 & 159.0 & 3.59$\pm$0.82 & \\
Dw 19 & 14:10:20.1 & +54:45:50 & 20.4$\pm$0.1 & 19.8$\pm$0.1 & 4.56$\pm$0.23 & 154.7$\pm$7.6 & 136.0 & 0.92$\pm$0.07 & Nucleated\\
Dw 20 & 14:10:01.6 & +54:25:11 & 21.0$\pm$0.1 & 20.6$\pm$0.1 & 3.98$\pm$0.46 & 135$\pm$16 & 121.6 & 1.04$\pm$0.16 & \\
Dw 21 & 14:07:56.5 & +54:56:03 & 21.9$\pm$0.2 & 20.8$\pm$0.1 & 3.26$\pm$0.74 & 111$\pm$25 & 110.0 & 0.77$\pm$0.16 & \\
\enddata
\tablenotetext{a}{From GALFIT model.}
\tablenotetext{b}{Assuming a distance to M101; $D$=7 Mpc.}
\tablenotetext{c}{Assuming a distance to NGC5485; $D$=27 Mpc.}
\end{deluxetable}

\floattable
\begin{deluxetable}{c|cc|c|c|c|c|c|c|c}
\tablenum{1}
\tablecaption{Dwarf Properties \label{tab:obj_prop_1}}
\tablehead{
\colhead{Name} & \colhead{RA} & \colhead{Dec} & \colhead{$g$ band} & \colhead{$r$ band} & \colhead{Half light} & \colhead{Half light} & \colhead{Projected} & \colhead{S\'{e}rsic} & \colhead{Notes}\\
\colhead{} & \colhead{} & \colhead{} & \colhead{Magnitude\tablenotemark{a}} & \colhead{Magnitude\tablenotemark{a}} & \colhead{radius\tablenotemark{a}} &\colhead{radius\tablenotemark{b}} & \colhead{distance from} &  \colhead{Index\tablenotemark{a}} & \colhead{(Confirmed}\\
\colhead{} & \colhead{} & \colhead{} & \colhead{} & \colhead{} & \colhead{(Arcsec)}& \colhead{(pc)} & \colhead{ M101\tablenotemark{b} (pc)}& \colhead{} & \colhead{associations)}}
\colnumbers
\startdata
Dw 22 & 14:03:03.3 & +54:47:12 & 21.2$\pm$0.1 & 20.8$\pm$0.1 & 3.41$\pm$0.31 & 116$\pm$11 & 53.6 & 0.91$\pm$0.11 & \\
Dw 23 & 14:07:08.4 & +54:33:49 & 22.1$\pm$0.3 & 20.6$\pm$0.4 & 9.3$\pm$7.6 & 310$\pm$260 & 74.6 & 2.06$\pm$0.56 & \\
Dw 24 & 14:06:48.0 & +54:23:36 & 21.6$\pm$0.1 & 20.8$\pm$0.1 & 2.67$\pm$0.21 & 90.5$\pm$7.3 & 64.1 & 0.77$\pm$0.08 & Nucleated\\
Dw 25 & 14:09:47.5 & +54:02:55 & 21.1$\pm$0.2 & 21.4$\pm$ 0.1& 6.0$\pm$2.2 & 204$\pm$76 & 123.2 & 1.60$\pm$0.37 & \\
Dw 26 & 14:08:50.4 & +53:27:24 & 20.2$\pm$0.1 & 19.8$\pm$0.1 & 5.08$\pm$0.10 & N/A & N/A & 0.58$\pm$0.03 & Nucleated\\
 &  &  &  &  & & & & & (Background)\\
Dw 27 & 14:12:22.9 & +54:31:00 & 21.2$\pm$0.1 & 20.6$\pm$0.1 & 3.78$\pm$0.17 & 128.4$\pm$5.8 & 164.3 & 0.85$\pm$0.09 & \\
Dw 28 & 14:08:46.3 & +55:05:02 & 20.7$\pm$0.1 & 20.3$\pm$0.1 & 4.18$\pm$0.23 & 141.7$\pm$7.7 & 133.0 & 0.91$\pm$0.08 & \\
Dw 29 & 14:08:09.9 & +55:04:33 & 22.0$\pm$0.2 & 21.5$\pm$0.1 & 3.38$\pm$0.80 & 115$\pm$27 & 124.7 & 0.50$\pm$0.25 & \\
Dw 30 & 14:08:08.7 & +53:35:45 & 21.0$\pm$0.1 & 20.5$\pm$0.1 & 5.80$\pm$0.29 & 196.9$\pm$9.9 & 127.8 & 0.70$\pm$0.08 & Nucleated\\
Dw 31 & 14:07:41.7 & +54:35:18 & 19.4$\pm$0.1 & 18.8$\pm$0.1 & 5.91$\pm$0.23 & 200.5$\pm$7.9 & 84.9 & 0.61$\pm$0.05 & \\
Dw 32 & 14:07:46.4 & +54:15:26 & 19.0$\pm$0.1 & 18.8$\pm$0.1 & 10.70$\pm$0.25 & 363.1$\pm$8.5 & 82.1 & 1.17$\pm$0.03 & \\
Dw 33 & 14:08:33.8 & +55:26:49 & 19.8$\pm$0.1 & 19.2$\pm$0.2 & 4.59$\pm$0.21 & 155.7$\pm$7.0 & 163.9 & 0.68$\pm$0.04 & \\
Dw 34 & 14:13:23.1 & +54:04:52 & 21.3$\pm$0.1 & 20.4$\pm$0.3 & 3.57$\pm$0.70 & 121$\pm$24 & 184.8 & 1.36$\pm$0.28 & \\
Dw 35 & 14:05:36.2 & +54:49:02 & 22.1$\pm$0.2 & 21.6$\pm$0.1 & 2.62$\pm$0.58 & 89$\pm$20 & 71.2 & 0.91$\pm$0.22 & \\
Dw 36 & 14:02:27.7 & +54:58:59 & 21.4$\pm$0.1 & 21.0$\pm$0.1 & 2.46$\pm$0.29 & 83.4$\pm$9.8 & 78.6 & 1.08$\pm$0.17 & \\
Dw 37 & 13:56:09.7 & +54:25:43 & 21.4$\pm$0.1 & 20.9$\pm$0.1 & 2.50$\pm$0.16 & 84.8$\pm$5.6 & 125.7 & 0.80$\pm$0.10 & \\
Dw 38 & 14:01:17.6 & +54:21:14 & 20.1$\pm$0.1 & 19.8$\pm$0.1 & 4.33$\pm$0.16 & 147.1$\pm$5.5 & 34.1 & 0.72$\pm$0.04 & \\
\enddata
\tablenotetext{a}{From GALFIT model.}
\tablenotetext{b}{Assuming a distance to M101; $D$=7 Mpc.}
\tablenotetext{c}{Assuming a distance to NGC5485; $D$=27 Mpc.}
\end{deluxetable}

\subsection{Distribution of New Dwarfs around M101}{\label{subsec:distrub}}

\begin{figure*}
 \begin{center}
 \includegraphics[width=18cm]{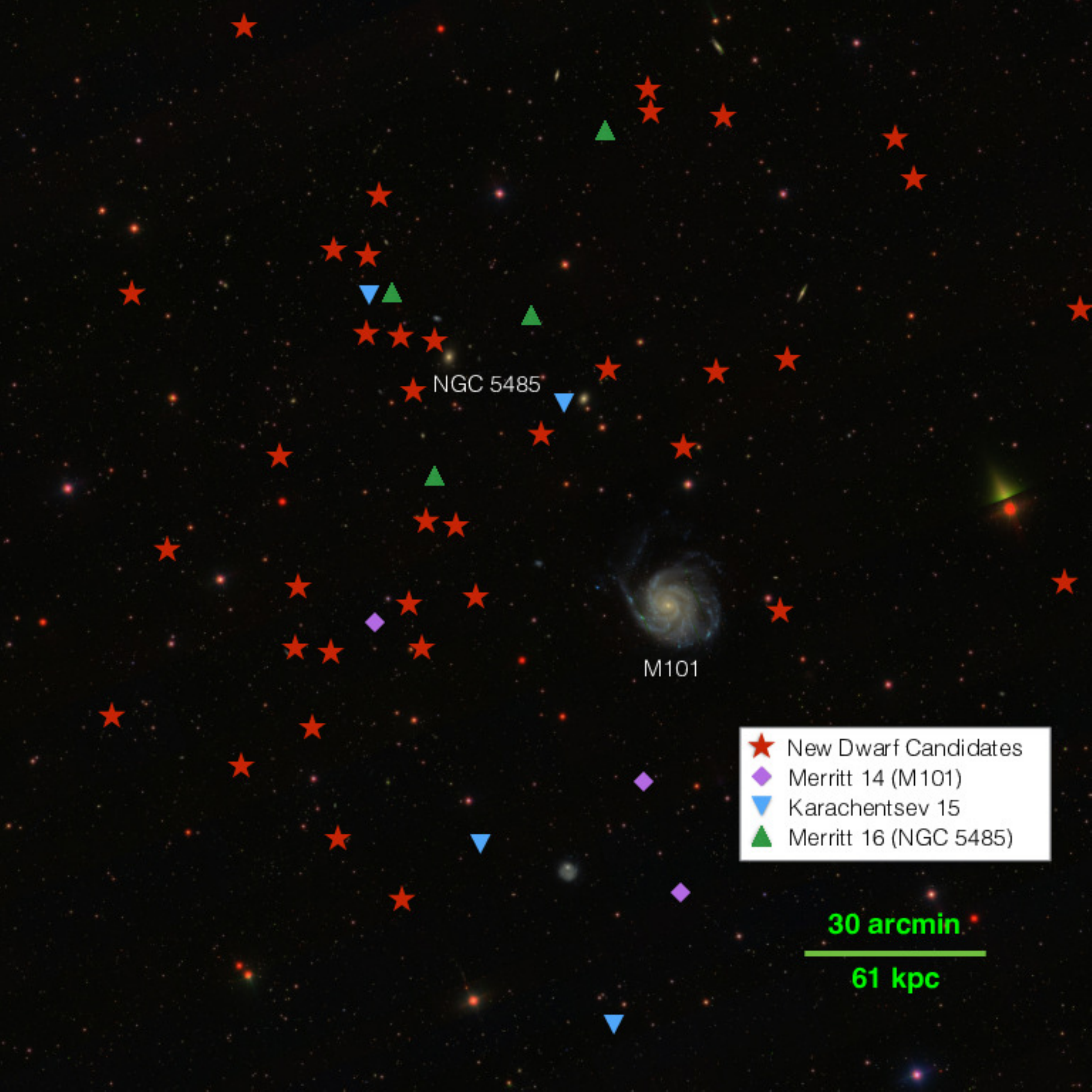}
 \caption{Unusual distribution of the newly discovered dwarfs around M101. Magenta diamonds are M14 dwarfs, green triangles are those initially reported in M14 but now believed to be members of the background NGC 5485 group in M16. Blue inverted triangles are K15 dwarfs. Red stars are our newly discovered dwarfs. Base image from SDSS. North is up, East is left. Image is 3x3 degrees. \label{fig:position}}
 \end{center}
\end{figure*}

The new dwarf population shows a distinct asymmetry as shown in Figure~\ref{fig:position}. Almost all of the candidates are grouped to the northeast of M101, with only two of the newly discovered dwarfs to the southwest. The dwarf candidate occurrence rate to the southwest is equivalent to $\sim$1 dwarf candidate per square degree, which is below the `background' dwarf candidate occurrence rate of $\sim$2 per square degree in the CFHTLS Wide Field 3 data set (Bennet et al. in prep). However if this low rate of dwarfs was replicated around the entire of M101 it would have $\sim$9 dwarfs total, which is only slightly poorer than the MW in this magnitude range. A similar asymmetry was reported in previous works (M14; K15; see also \citealt{muller17} for a possible large-scale plane of galaxies around the M101 group).

The grouping of dwarfs on the northeast side of M101 can be explained by the presence of the background NGC 5485 group (M16). If many of the proposed dwarfs are associated with this group, it would explain the large asymmetry seen in the candidate population. This background group contains a total of 6 bright (M$_B$$\leq$-14) members \citep{makarov14}. Assuming that the NGC 5485 group has a luminosity function between those of the MW/M31 \citep{mcconnachie12} and that of the Virgo cluster \citep{ferrarese16}, we would expect between $\sim$10 and $\sim$30 dwarf group members in the examined magnitude range, -13 $\lesssim$ M$_g$ $\lesssim$ -10. However concluding that all or most of our candidates are members of the background NGC 5485 group would imply that M101 (a MW analog in certain respects, but deficient in stellar halo; \citealt{vanDokkum14}) has a dwarf population sparser than those of the MW or M31. Recent work has suggested that the scatter in satellite numbers around MW-analogs is significant \citep{geha17}, and M101 may be a sparse member of the distribution.

An infalling group could be another explanation for the asymmetric distribution of dwarfs. 
Groups of dwarfs are commonly seen in simulations (e.g. \citealt{donghia08}) and observations (e.g. \citealt{tully06}; \citealt{stierwalt17}), and in this case would point to a MW analog accreting a group of dwarfs.
This idea is reinforced by the highly asymmetric HI and optical disk of M101 (\citealt{mihos12}, \citealt{mihos13}), which show features extending away from the disk to the east and northeast. 
An infalling group would seem be in contrast to the apparent inactive accretion history of the M101 group, as shown by the extremely small stellar mass fraction reported in its stellar halo \citep{vanDokkum14}. We expect that galactic halos formed from the remains of previous accretion events between a galaxy and its satellite system (e.g., \citealt{cooper13}) and the small mass fraction strongly implies that these events have previously been rare. If this were an infalling dwarf group with no massive galaxies we would expect these dwarfs to be star-forming. Also this explanation leaves the M101 group with the absence of an existing dwarf population.

\section{Conclusions} \label{sec:conclusion}

In this paper we have presented the development and application of a new algorithm designed to detect LSB galaxies. The creation and testing of this algorithm are described in detail in \hyperref[sec:analysis]{$\S$3}. With some additional refinement as discussed in that section, the algorithm can be used to search large amounts of data for unresolved LSB galaxies. 
This is particularly useful for identifying LSB objects that are not in groups with bright host galaxies allowing future searches of large scale surveys, both in the field and galaxy clusters, to be searched faster for LSB objects with well-characterized completeness limits.

The algorithm has been tested on a 9 deg$^2$ region roughly centered on M101, extending the search $\sim$1.2 magnitudes deeper than previous work. 
We have discovered 38 previously unreported objects and confirmed 11 previously reported LSB objects in this region (K15, M14). 
These new objects have apparent magnitudes in the range 19.0 $\leq$ m$_g$ $\leq$ 22.2 and half light radii in the range 3-16''. These properties are consistent with Local Group dwarf galaxies at M101 distance ($\sim$7 Mpc). At this distance, they are projected within the virial radius ($\sim$260 kpc). Their association with a massive host is supported by the lack of NUV emission, as dwarfs within the virial radius of a large galaxy should be stripped of gas and have no ongoing star formation. 

No new UDGs were found in the M101 group, despite sensitivity to this area of parameter space. This is in line with expectations, which are that a group this small should not have a UDG population (\citealt{roman17}, \citealt{vanderBurg17}). Seven of the candidates would be UDGs if at the distance of the background NGC 5485 group, this is also consistent with the UDG number to group size relation which predicts $\sim$5 UDGs in a group this size.

The positions of these new discoveries show a skewed distribution, with almost all objects to the northeast and only two candidates to the southwest of M101 (see Figure~\ref{fig:position}). This asymmetry can be explained either by the presence of the background NGC 5485 group (which is to the northeast of M101), or if the new candidates are part of an infalling dwarf group. However, these explanations do not explain the lack of an existing dwarf population around M101, whose numbers could be on the low end of that seen among MW analogs (e.g. \citealt{geha17}).  We will explore this further once the membership status of our sample is in hand. 

We have HST imaging and HI follow up scheduled for several of the dwarf candidates to study the environment of these objects. The HI observations (Karunakaran et al. in prep) will potentially confirm velocity with a detection, or indicate that these objects are associated with a larger galaxy by a lack of HI emission. With distances derived from HST imaging using the Tip of the Red Giant Branch method, it should be possible to confirm or exclude an association with M101; a lack of a detected resolved stellar population would also exclude such an association (see D17). The combination of these observations should be able to tell if objects are background or part of the M101 or NGC 5485 groups. 

In the future we will be applying the detection algorithm from this work to publicly available wide-field datasets to understand the properties of diffuse dwarfs in a range of environments.

\acknowledgments
We are grateful to P.J. Marshall for initial discussions about the dwarf detection algorithm employed in this work.

 DJS acknowledges support from NSF grant AST-1412504.

Based on observations obtained with MegaPrime/MegaCam, a joint project of CFHT and CEA/IRFU, at the Canada-France-Hawaii Telescope (CFHT) which is operated by the National Research Council (NRC) of Canada, the Institut National des Science de l'Univers of the Centre National de la Recherche Scientifique (CNRS) of France, and the University of Hawaii. This work is based in part on data products produced at Terapix available at the Canadian Astronomy Data Centre as part of the Canada-France-Hawaii Telescope Legacy Survey, a collaborative project of NRC and CNRS.

This research made use of Astropy, a community-developed core Python package for Astronomy (Astropy Collaboration, 2013).

\vspace{5mm}
\facilities{CFHTLS, GALEX}

\software{Python, SExtractor, GALFIT}

\appendix

\section{Dwarf images}\label{sec:images}

Here we present the g-band images and GALFIT models of our new dwarf candidates (see Table \hyperref[tab:obj_prop_1]{1}). For each image the left panel is the g-band image, the center panel is the GALFIT model and right panel is the residuals after subtraction. North is up and East is left for all images. They are presented in numerical order, with Dw 1 first and Dw 38 last.

\begin{figure}
\figurenum{A}
\gridline{
          \fig{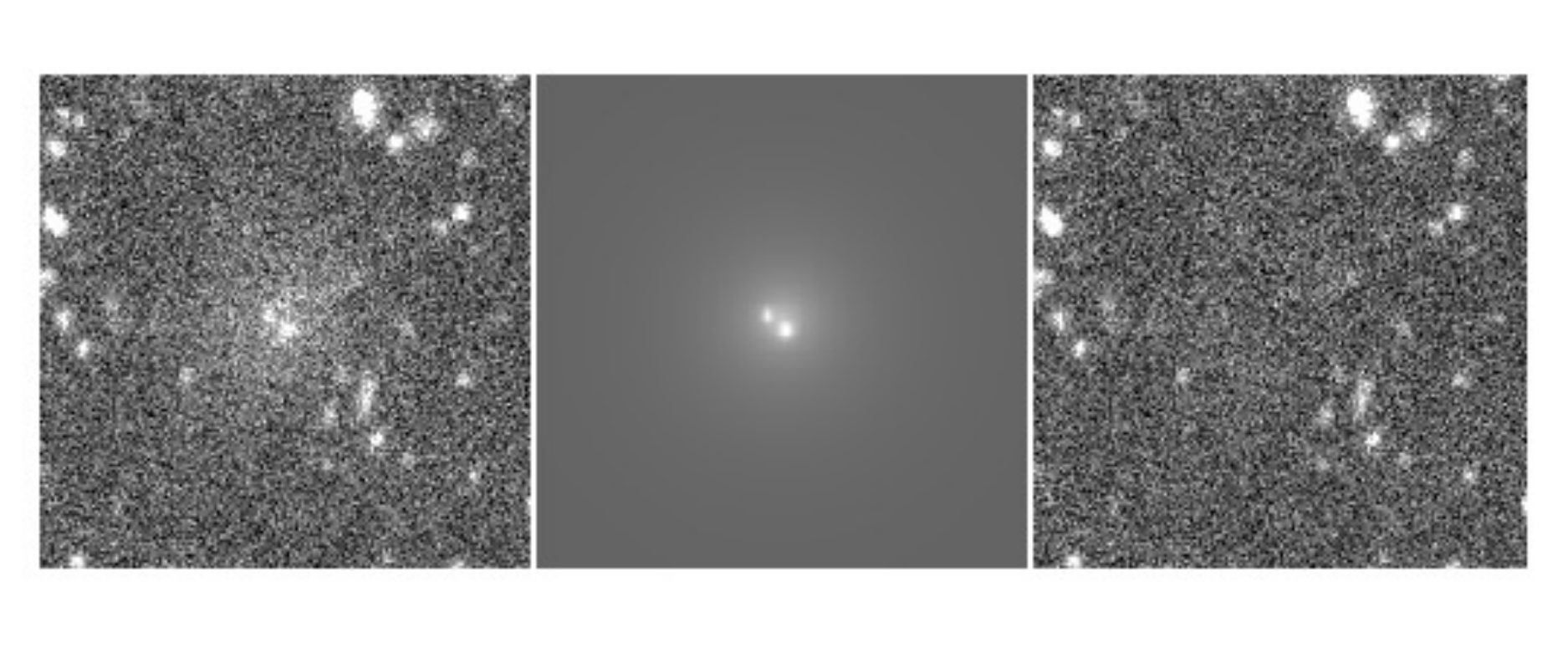}{0.33\textwidth}{(1)}
          \fig{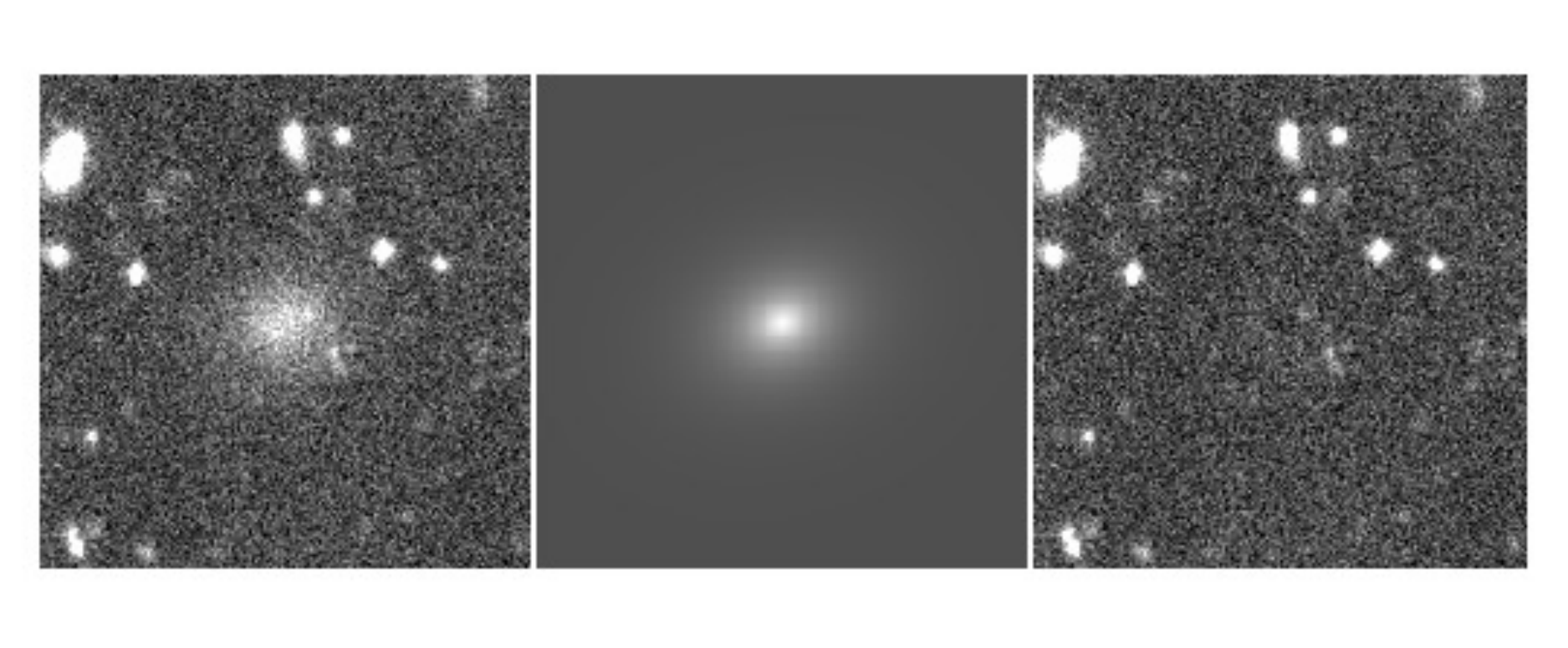}{0.33\textwidth}{(2)}
          \fig{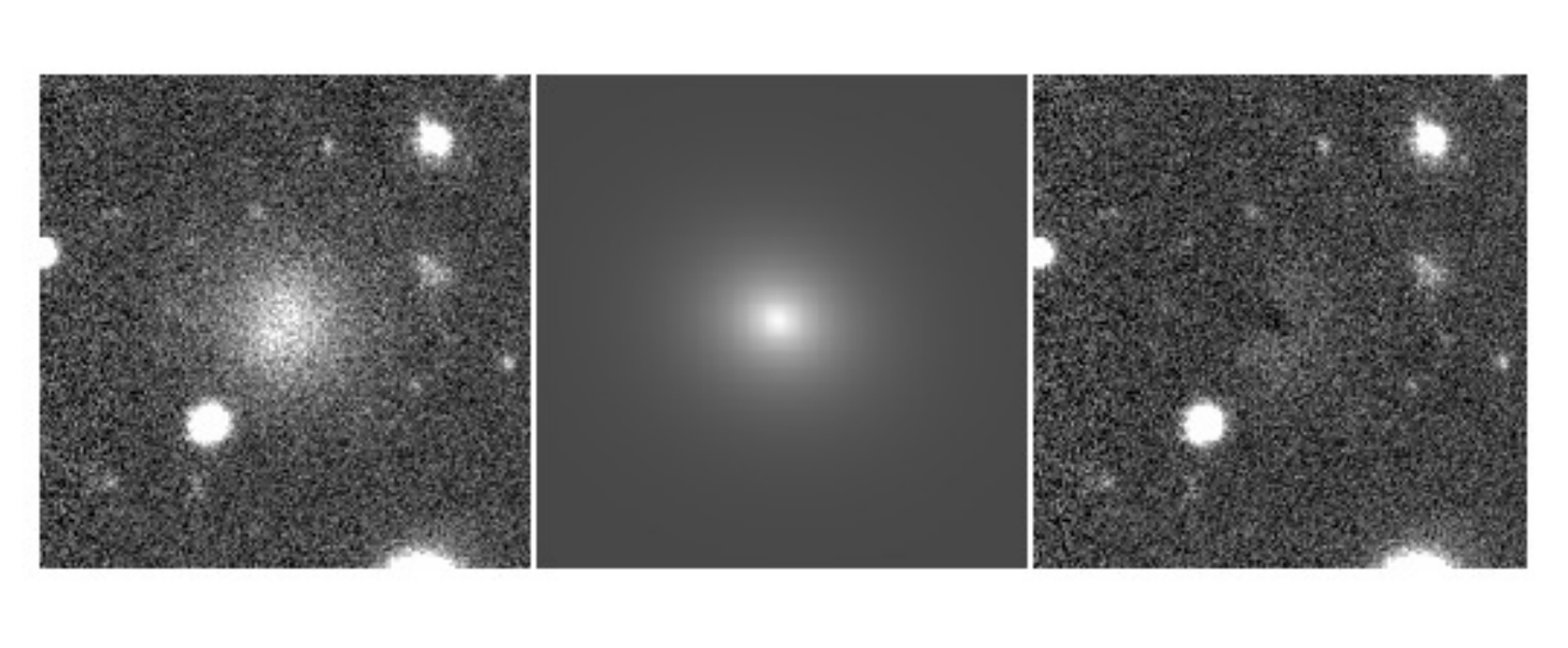}{0.33\textwidth}{(3)}
          }
\gridline{
          \fig{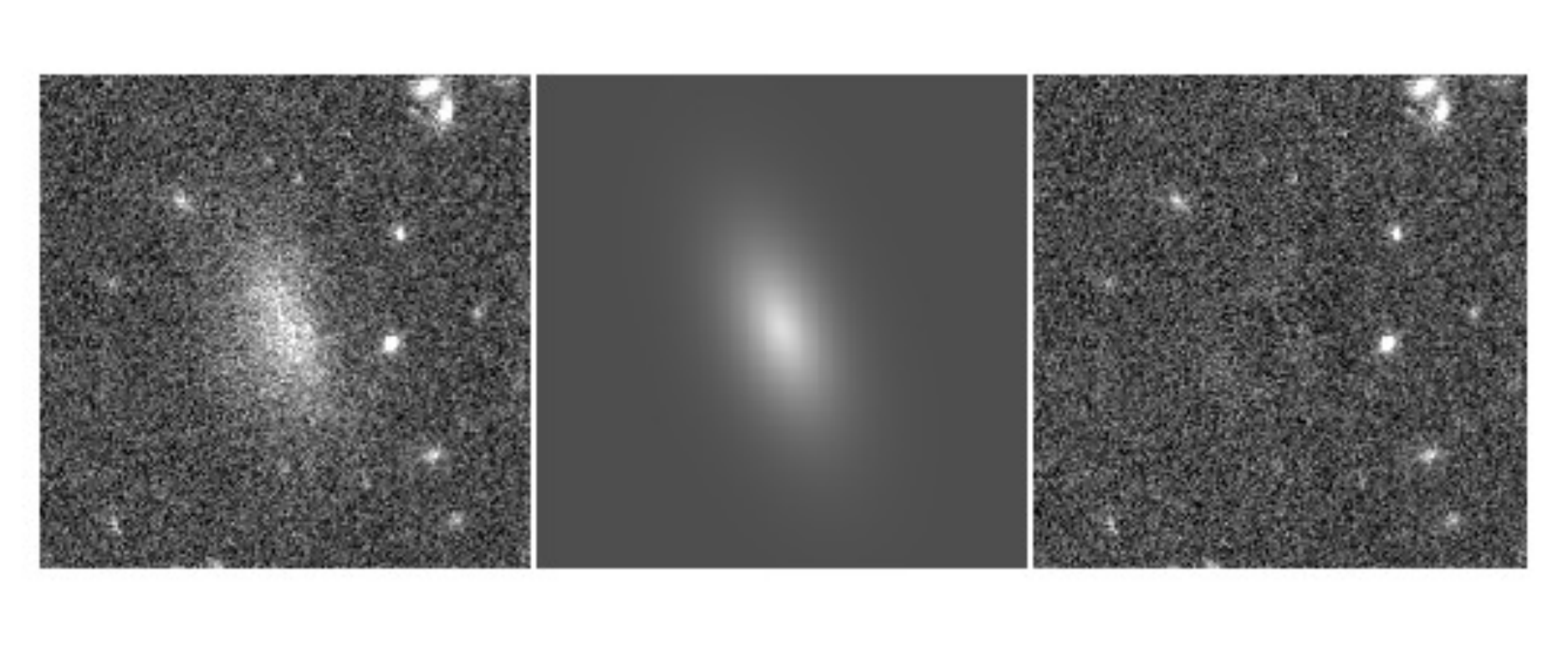}{0.33\textwidth}{(4)}
          \fig{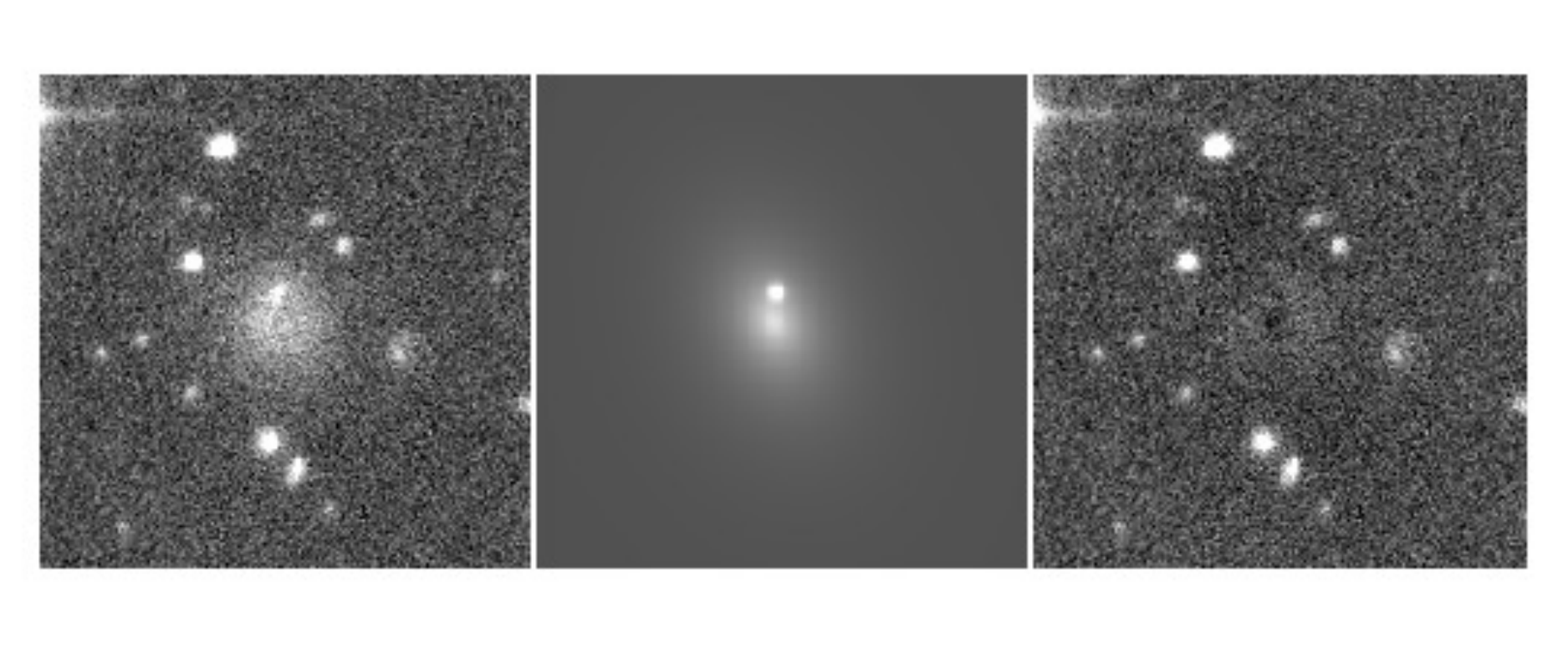}{0.33\textwidth}{(5)}
          \fig{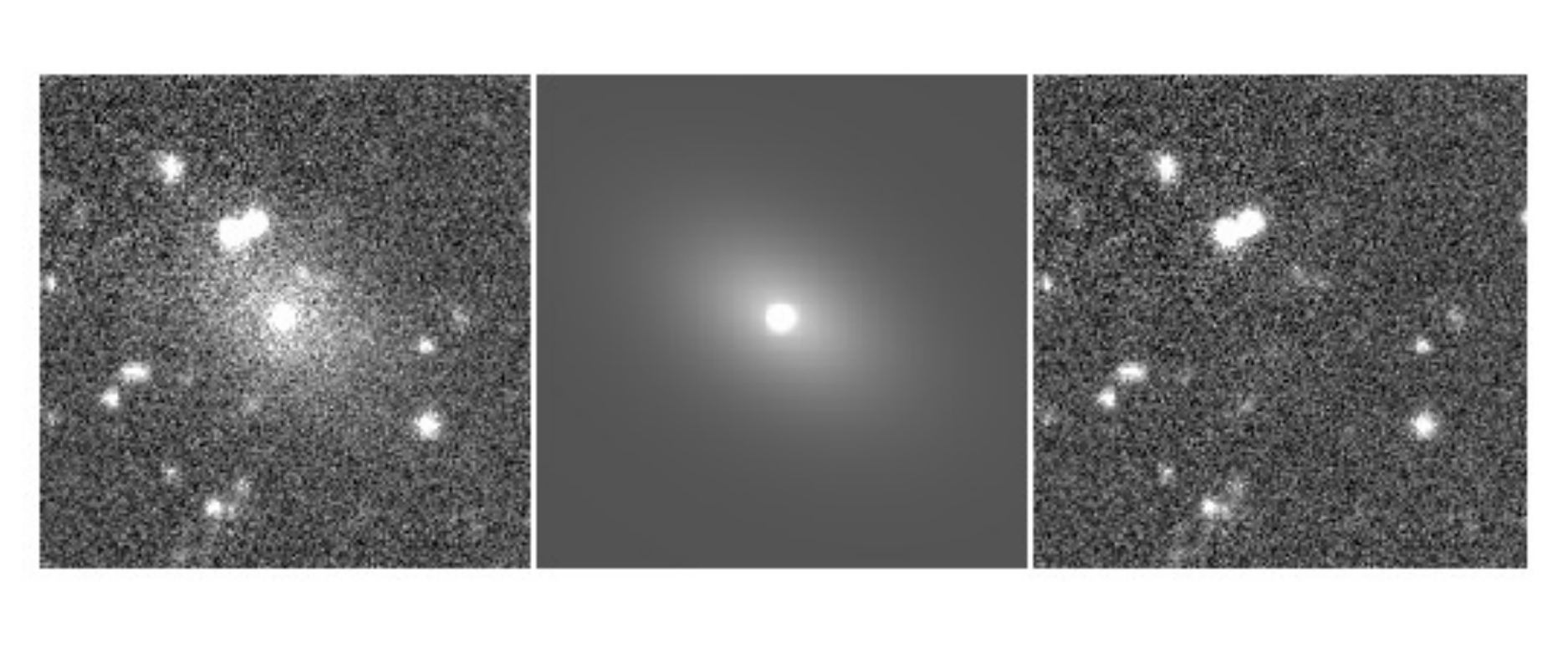}{0.33\textwidth}{(6)}
          }
 \gridline{
          \fig{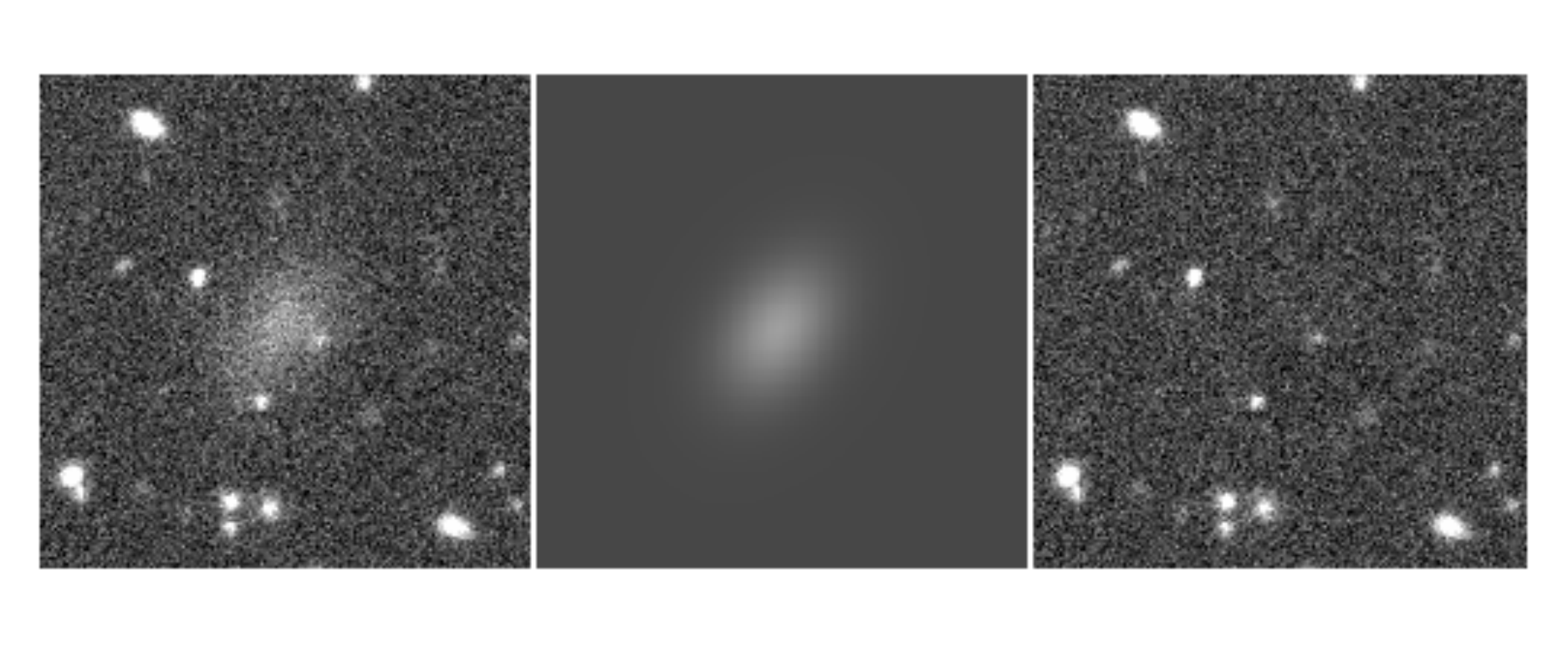}{0.33\textwidth}{(7)}
          \fig{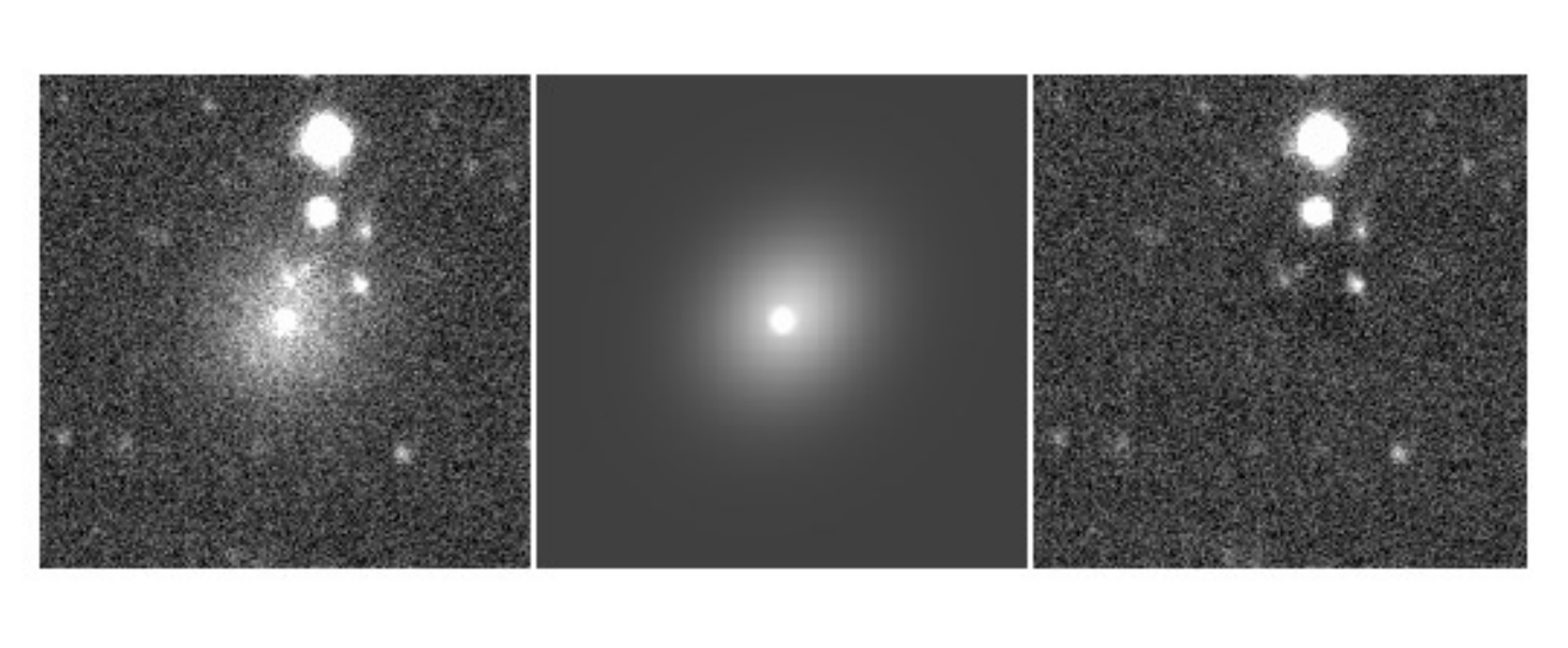}{0.33\textwidth}{(8)}
          \fig{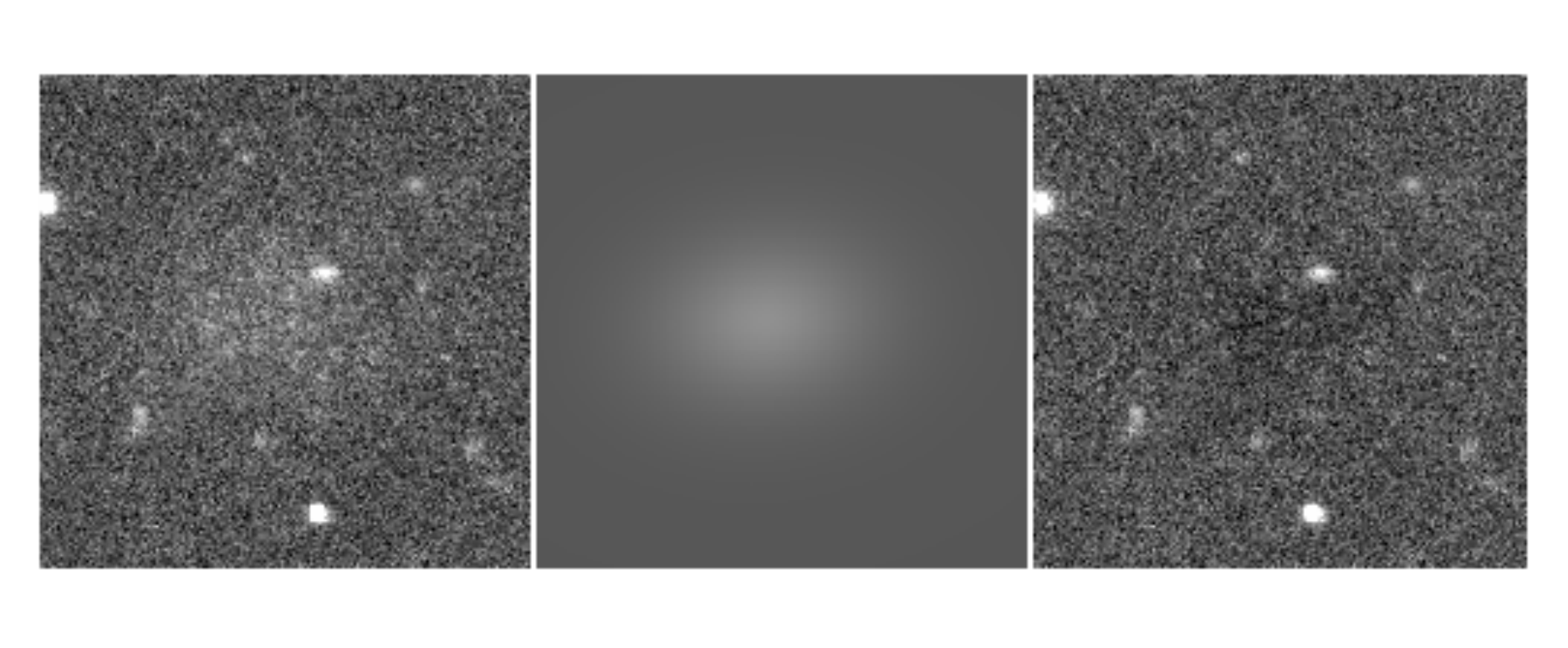}{0.33\textwidth}{(9)}
          }
\gridline{
          \fig{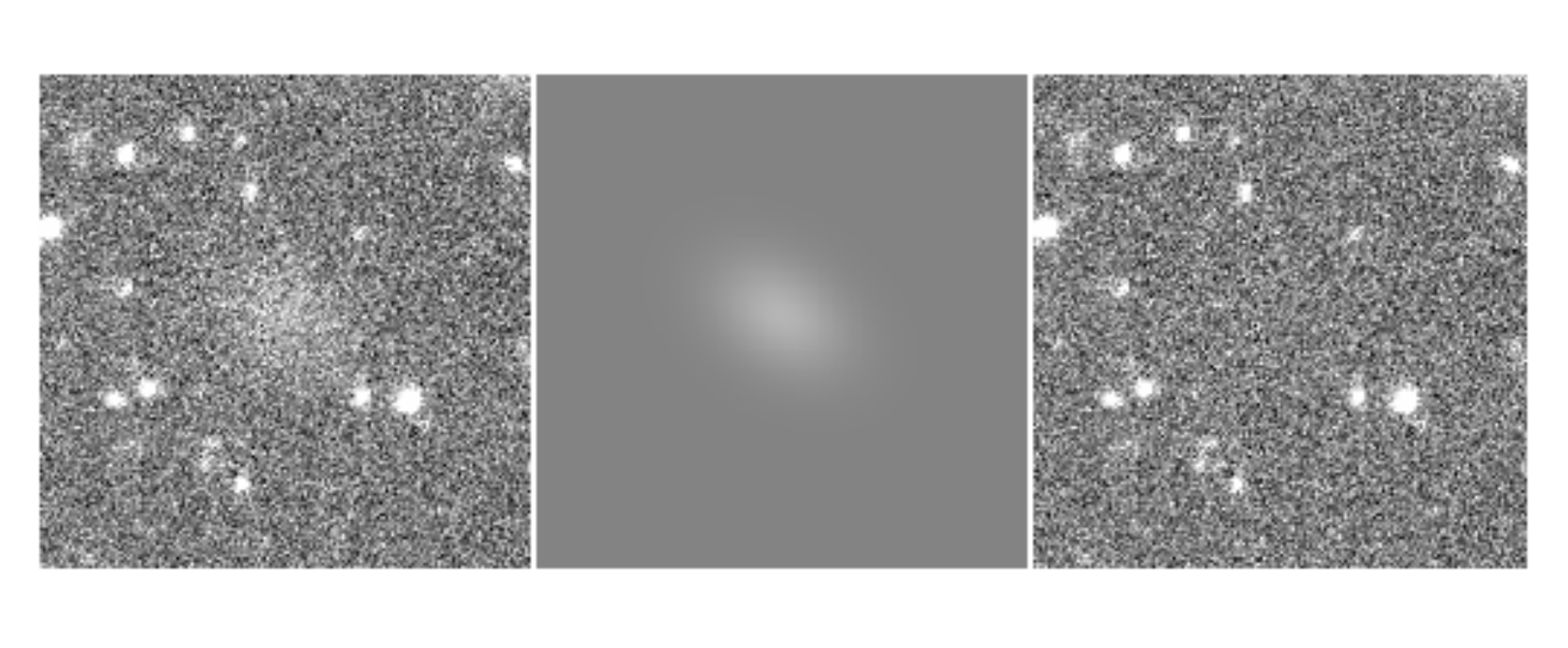}{0.33\textwidth}{(10)}
          \fig{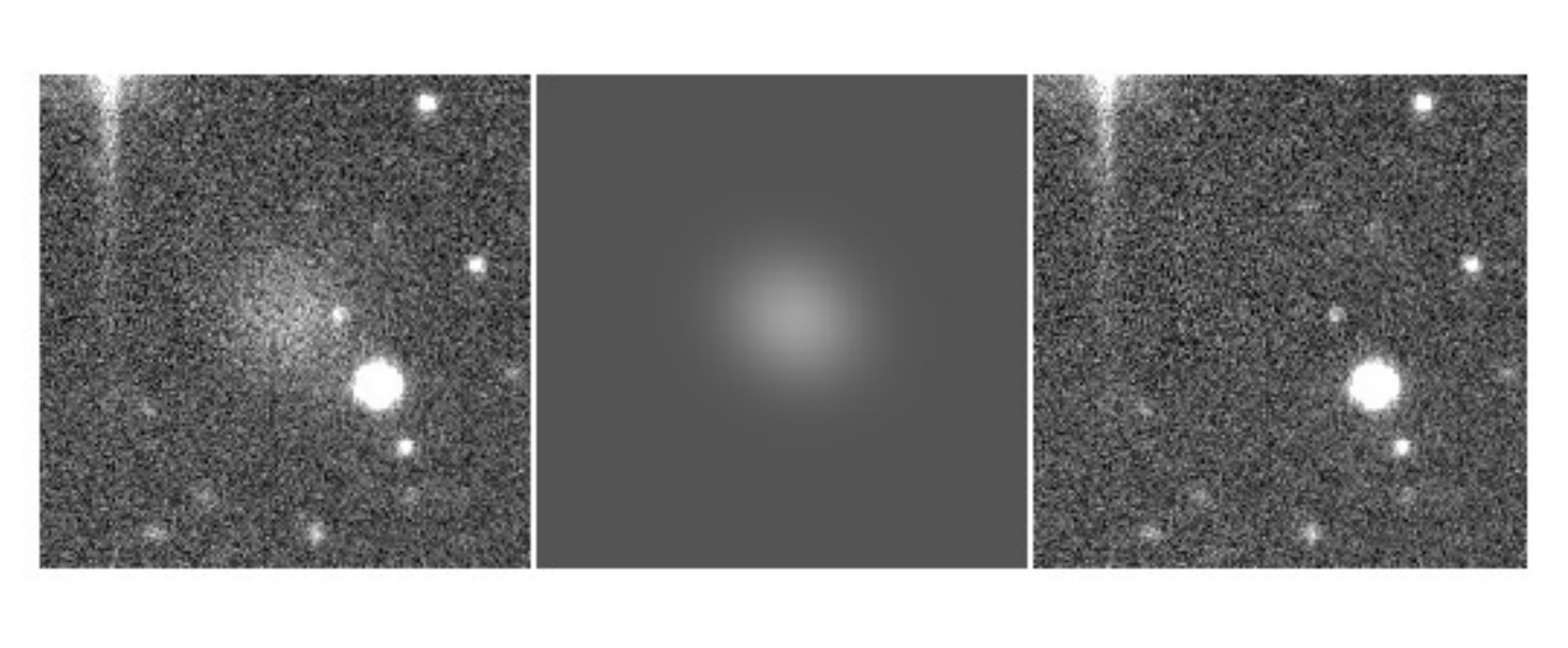}{0.33\textwidth}{(11)}
          \fig{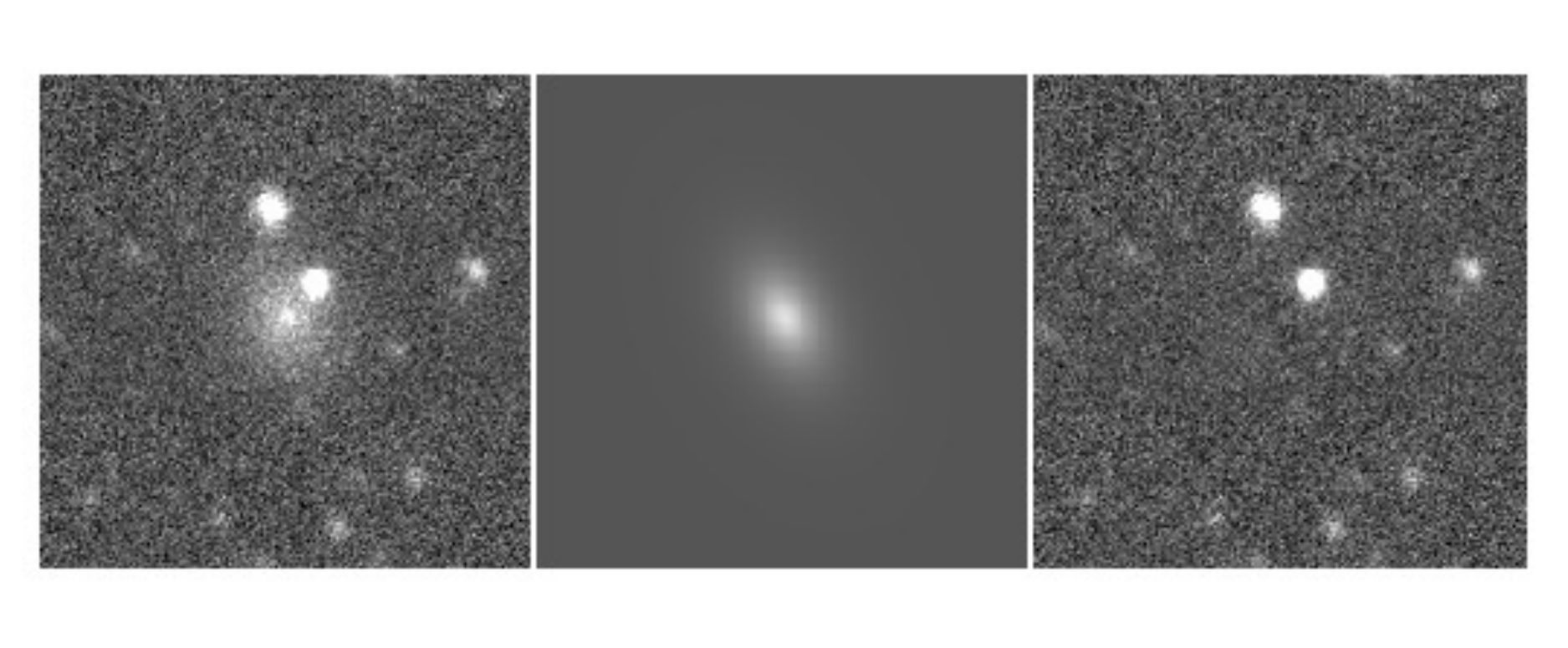}{0.33\textwidth}{(12)}
          }
\gridline{
          \fig{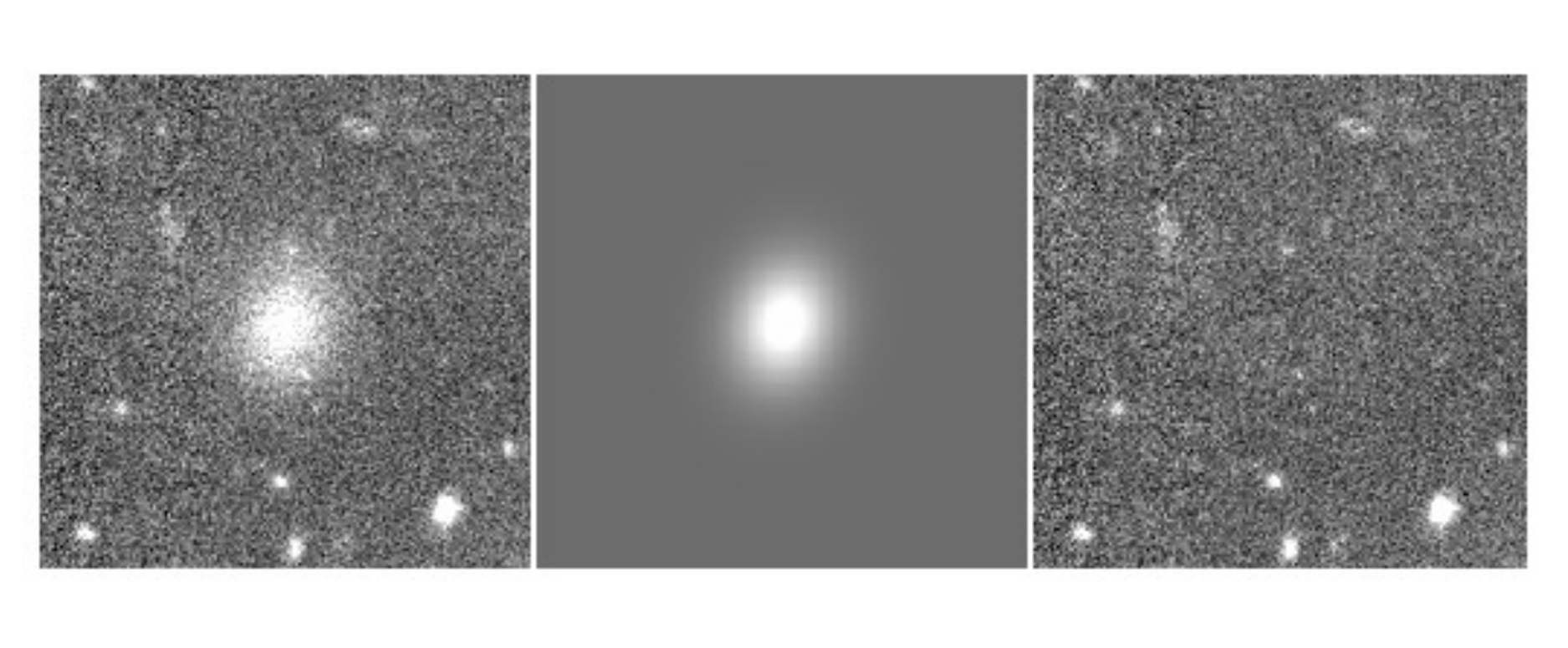}{0.33\textwidth}{(13)}
          \fig{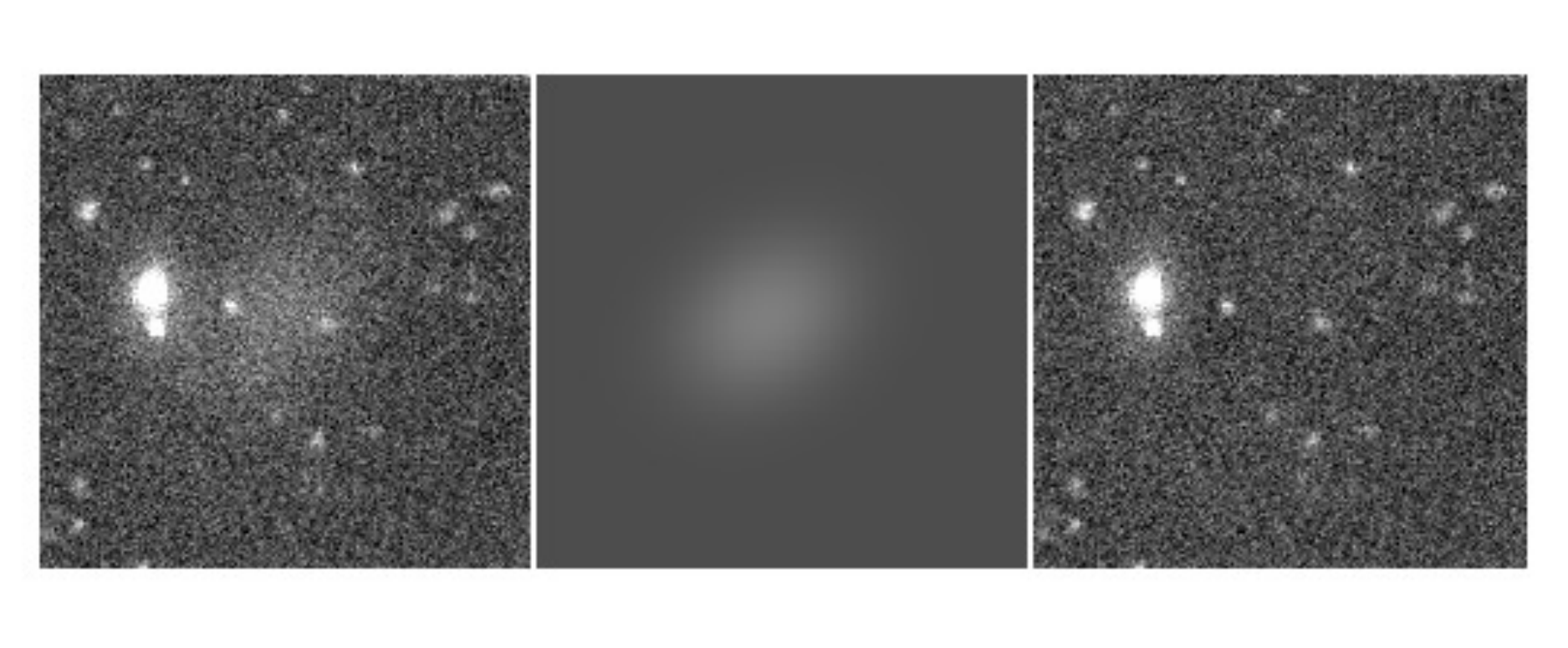}{0.33\textwidth}{(14)}
          \fig{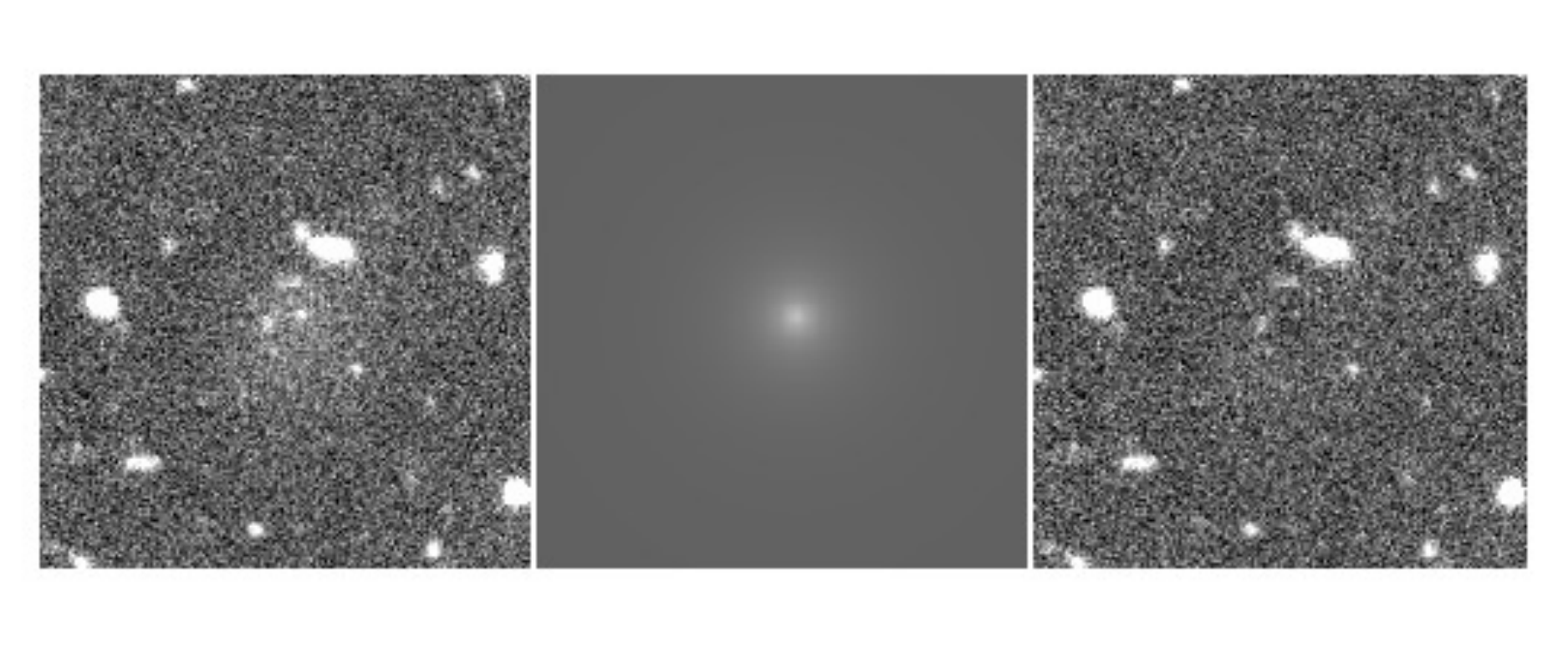}{0.33\textwidth}{(15)}
          }
\gridline{
          \fig{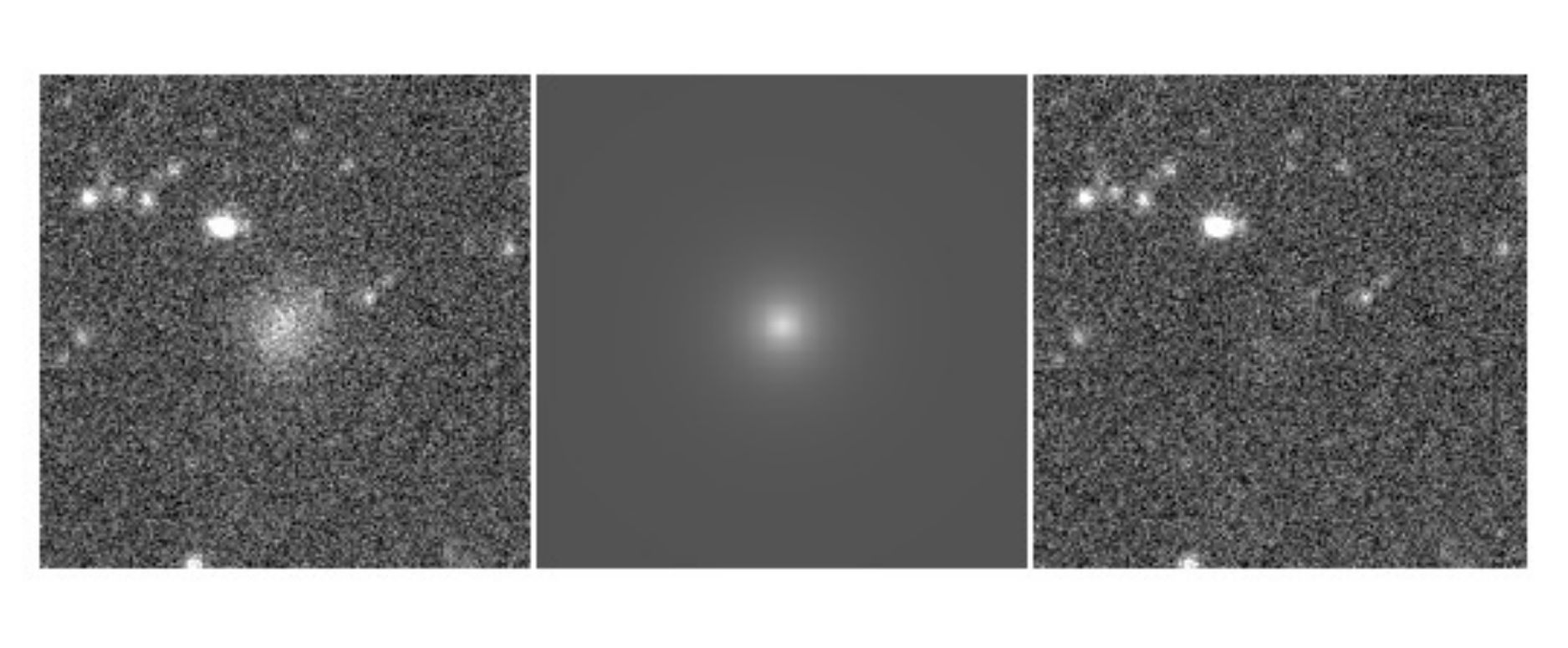}{0.33\textwidth}{(16)}
          \fig{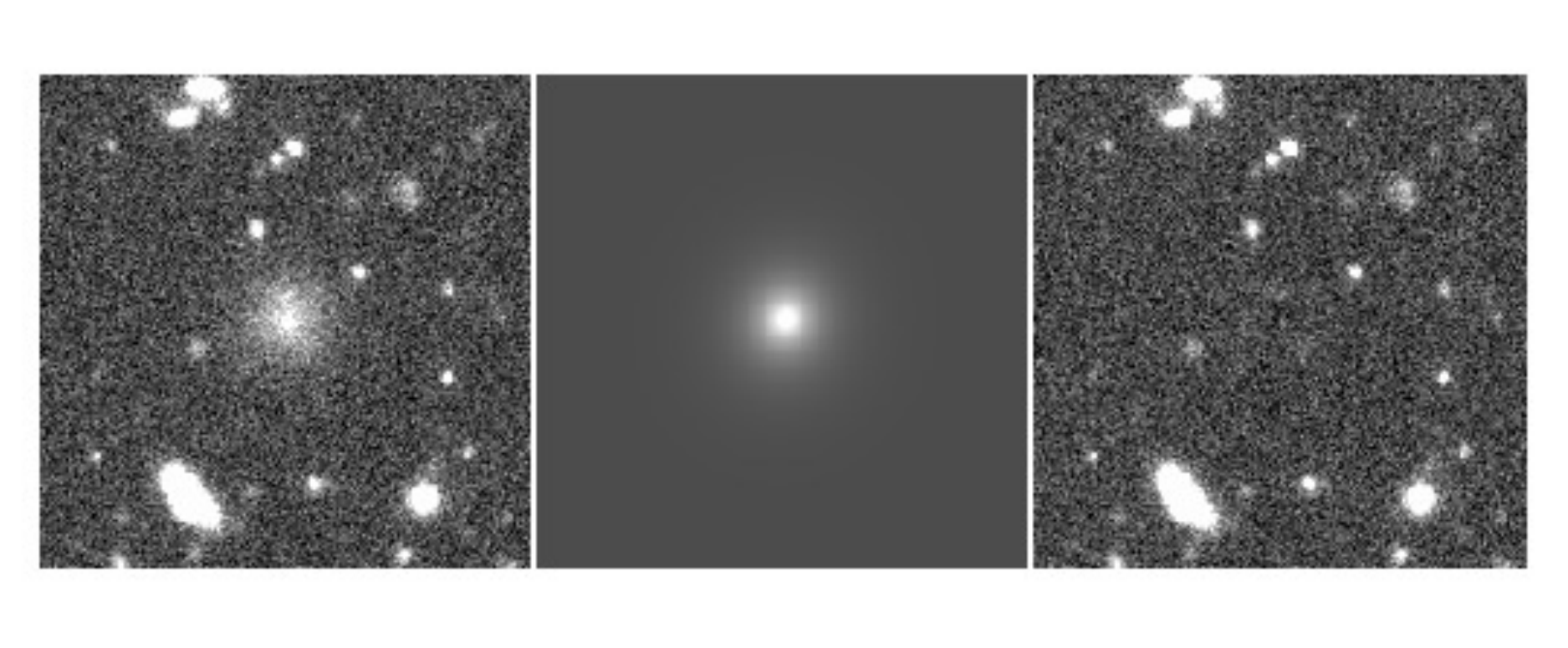}{0.33\textwidth}{(17)}
          \fig{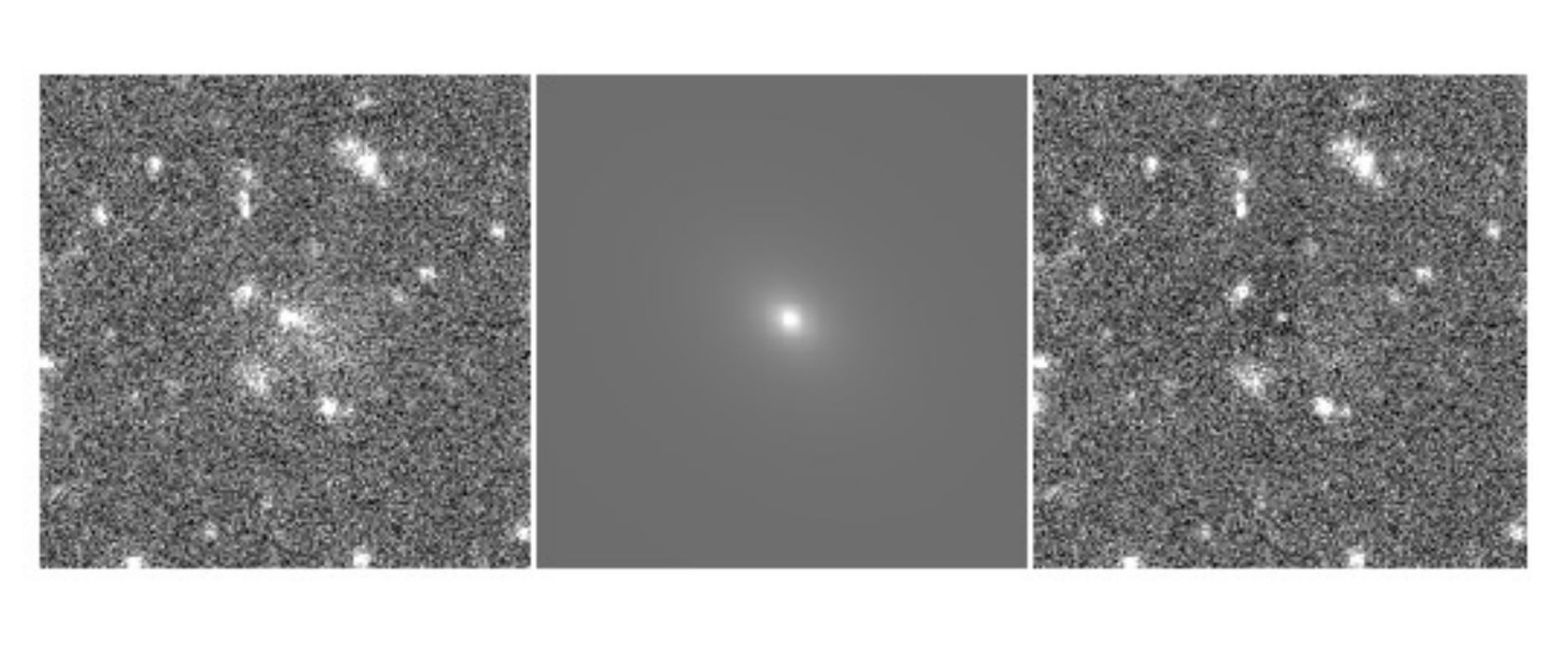}{0.33\textwidth}{(18)}
          }
\gridline{}
\caption{The results of our GALFIT fitting to the new dwarf galaxies. Images are 37.2''x32.7''. North is up, East is Left. Left panel: $g$ band CFHTLS image. Center panel: GALFIT model. Right panel: Residuals.\label{fig:galfit_1}}
\end{figure}

\begin{figure}
\figurenum{A cont}

\gridline{
          \fig{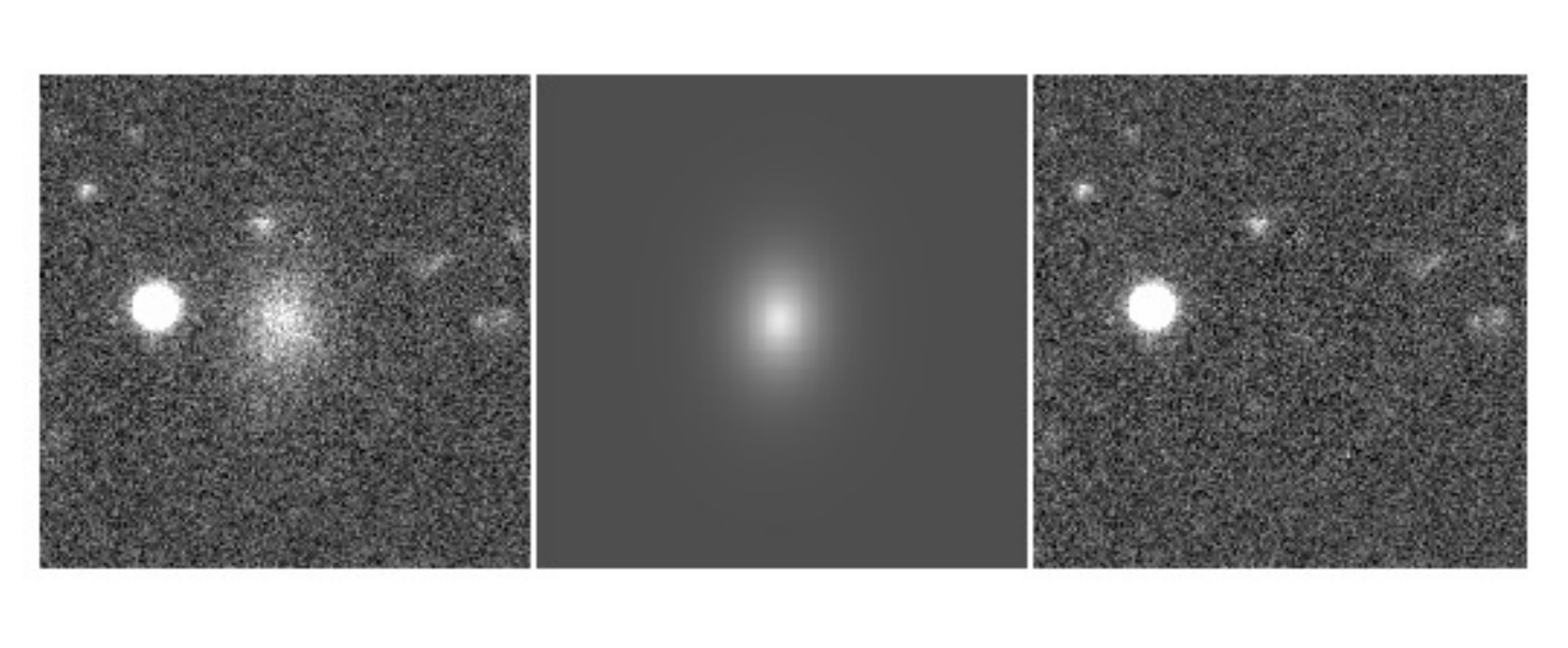}{0.33\textwidth}{(19)}
          \fig{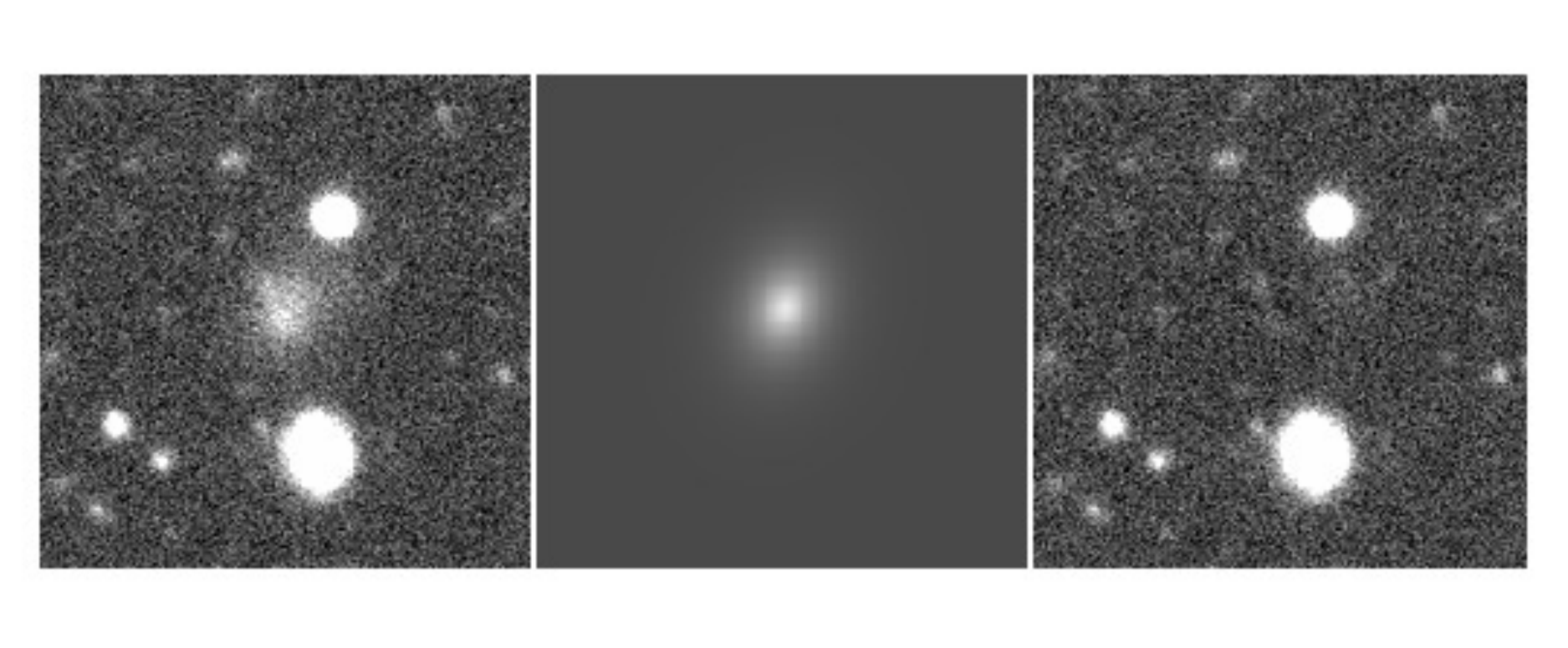}{0.33\textwidth}{(20)}
          \fig{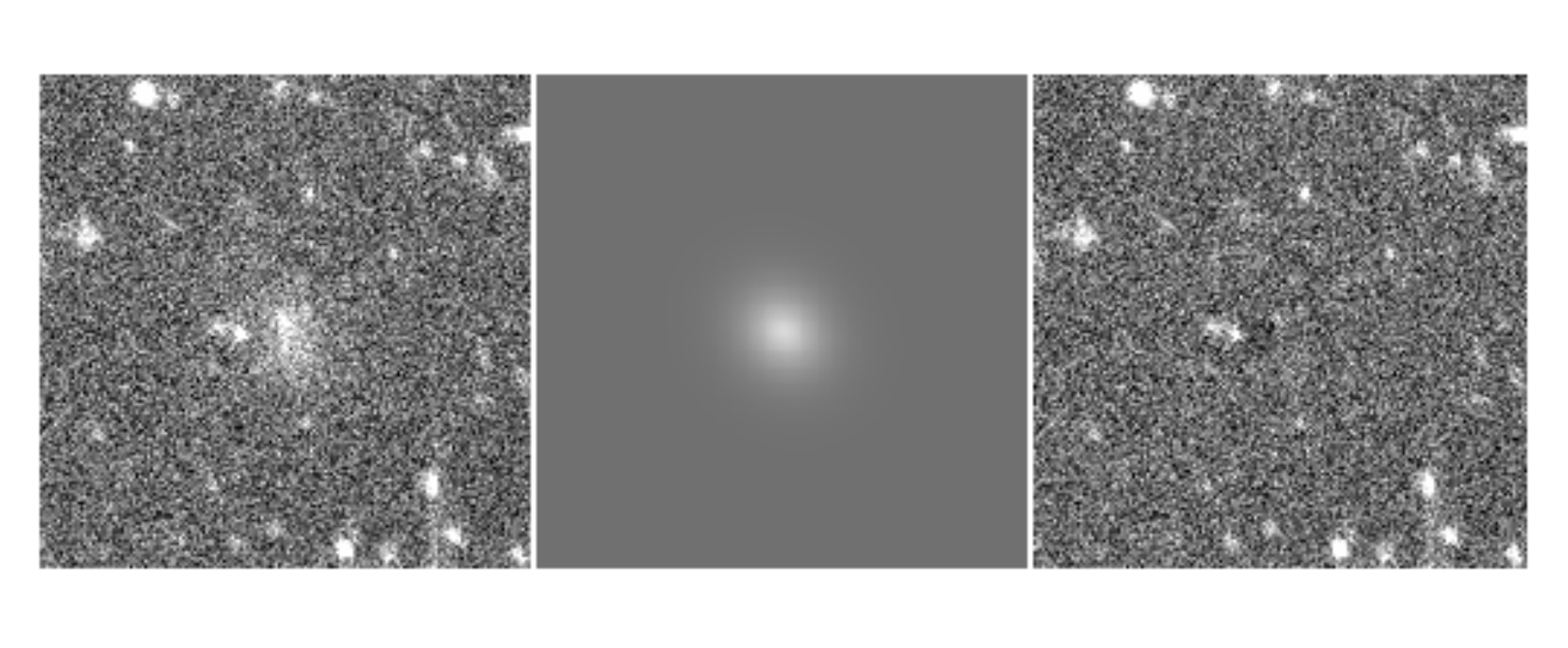}{0.33\textwidth}{(21)}
          }
\gridline{
          \fig{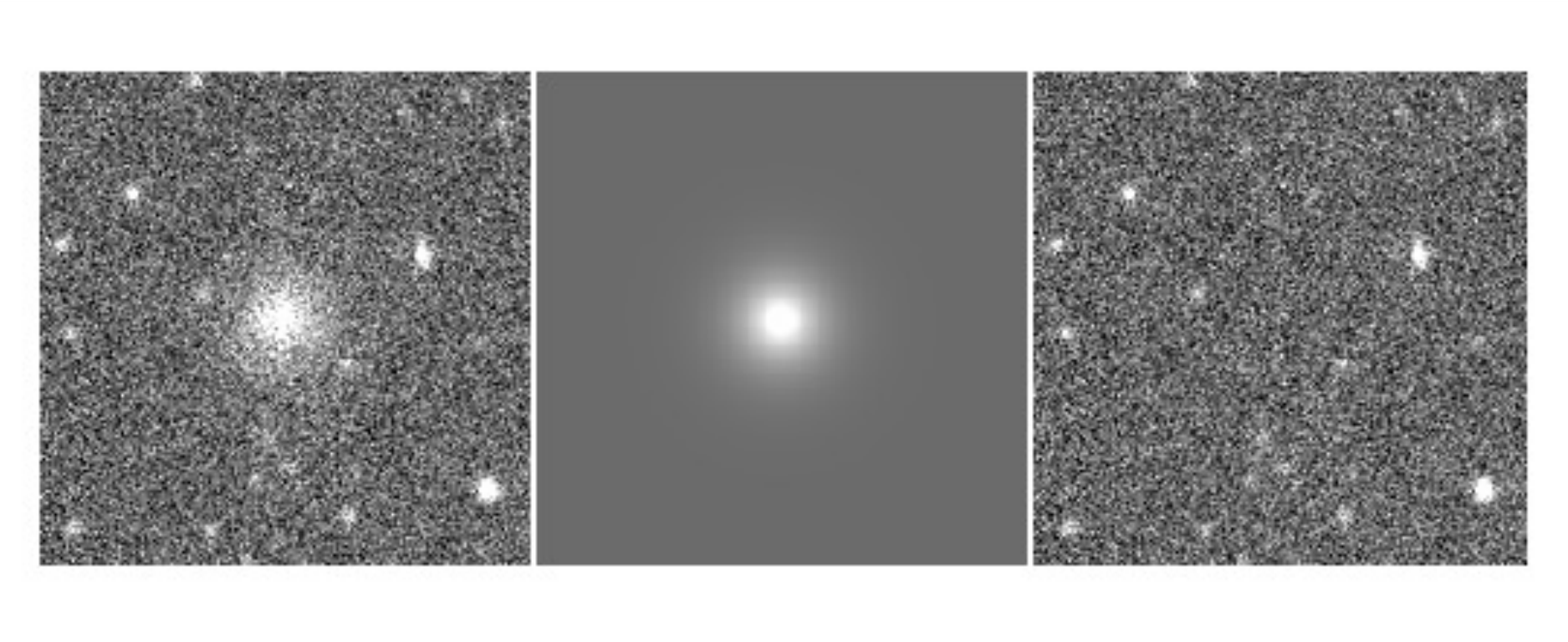}{0.33\textwidth}{(22)}
          \fig{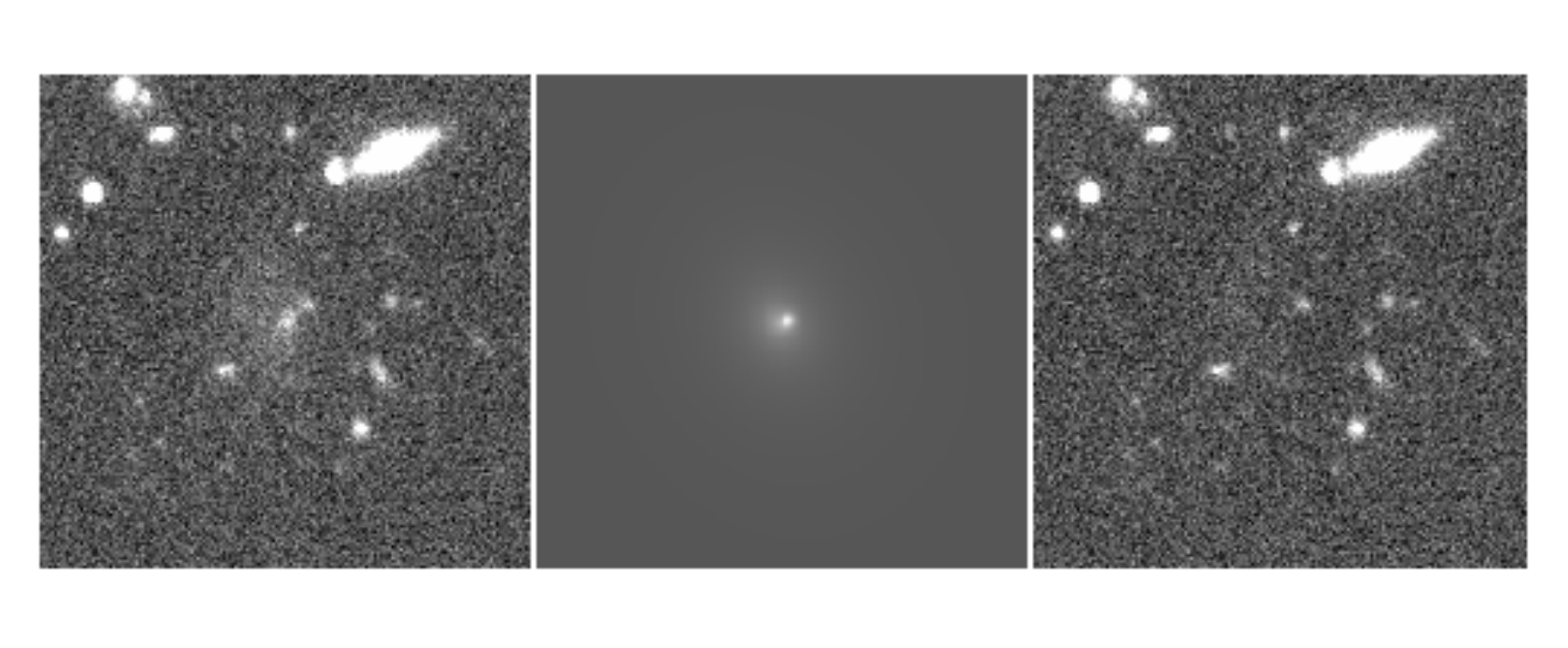}{0.33\textwidth}{(23)}
          \fig{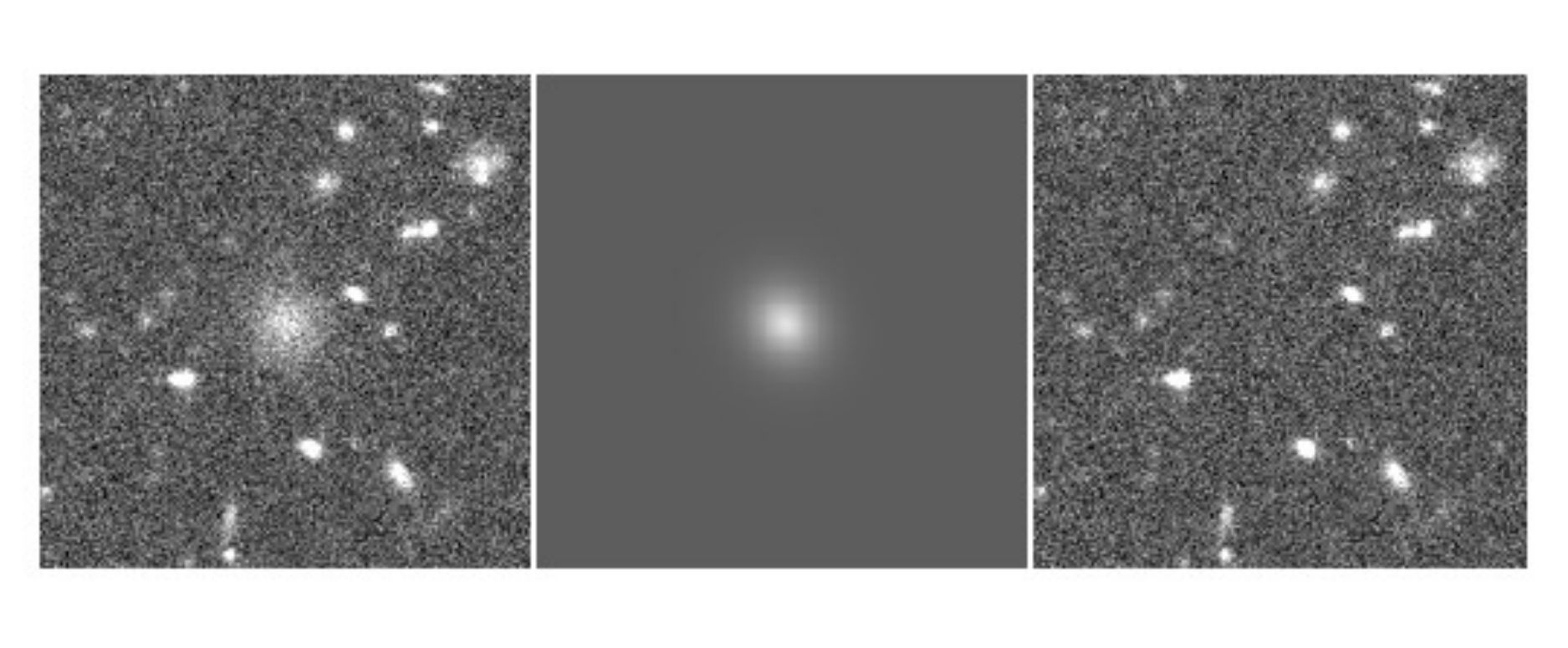}{0.33\textwidth}{(24)}
          }
\gridline{
          \fig{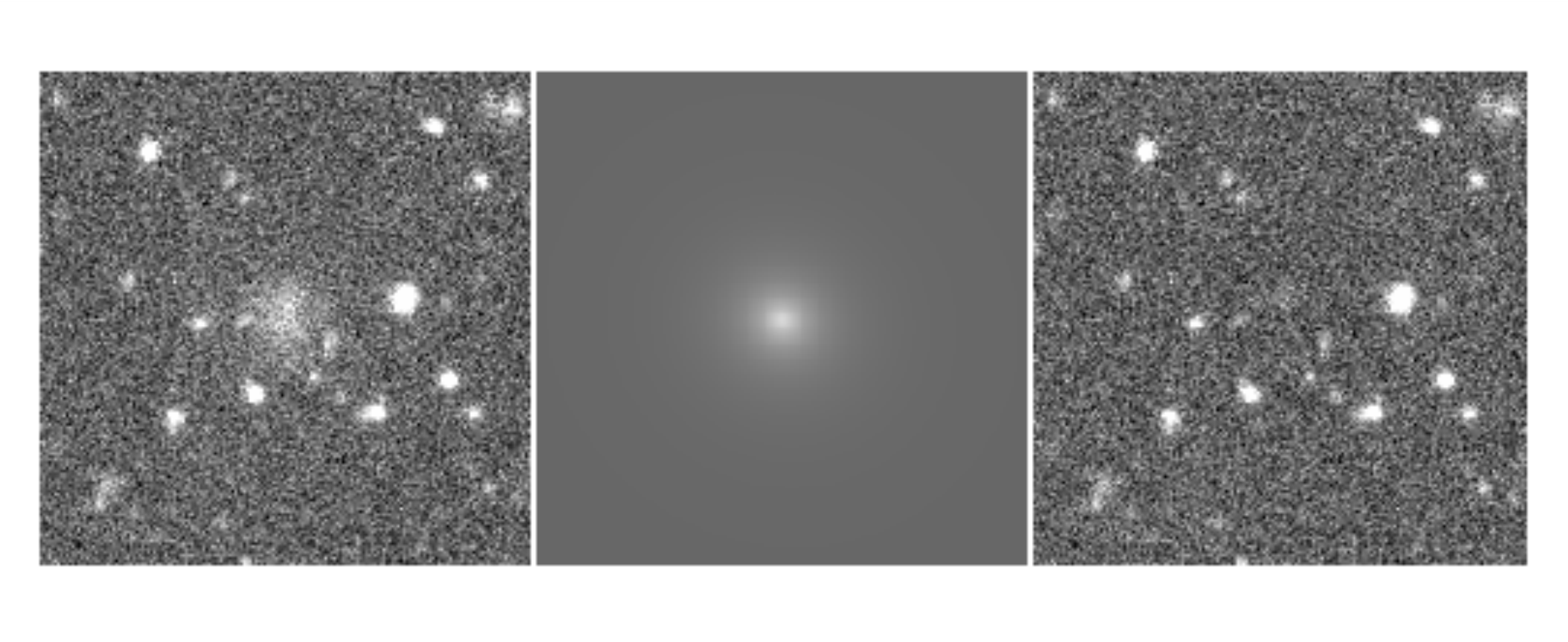}{0.33\textwidth}{(25)}
          \fig{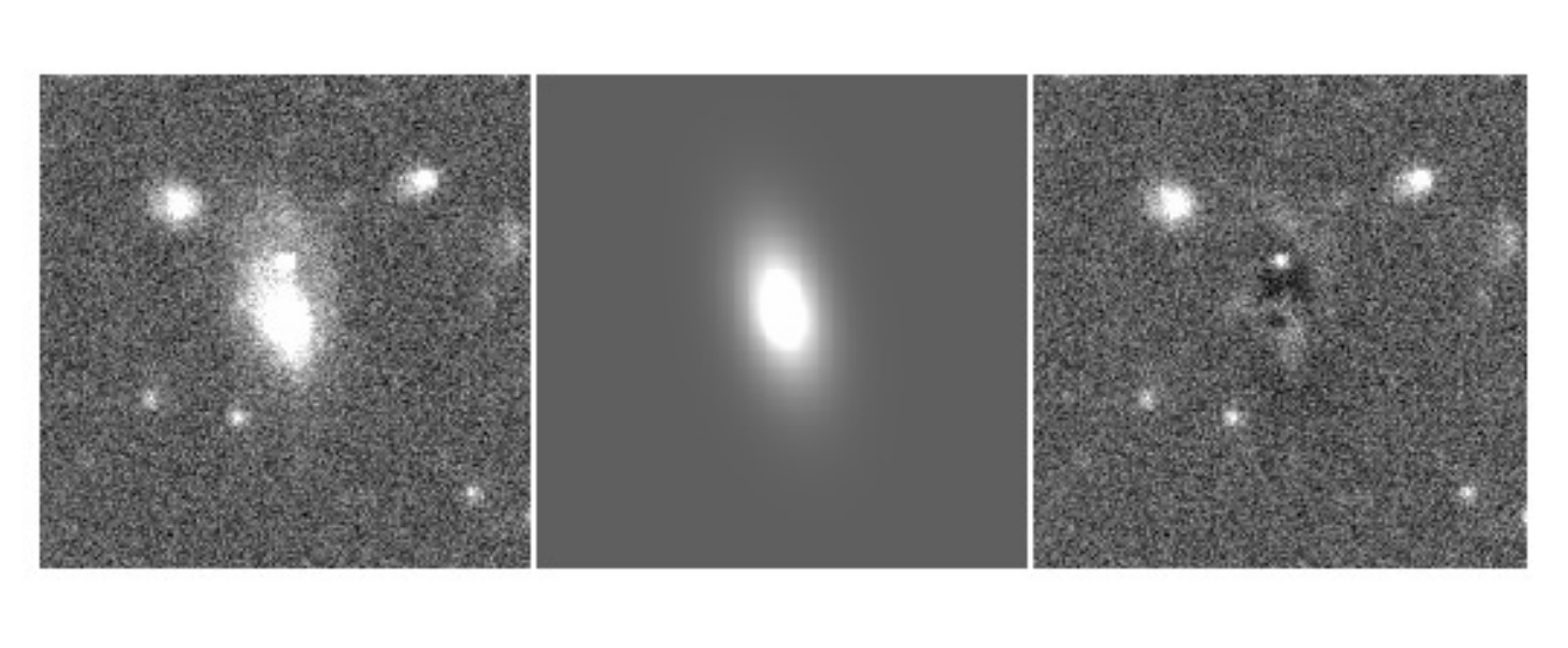}{0.33\textwidth}{(26)}
          \fig{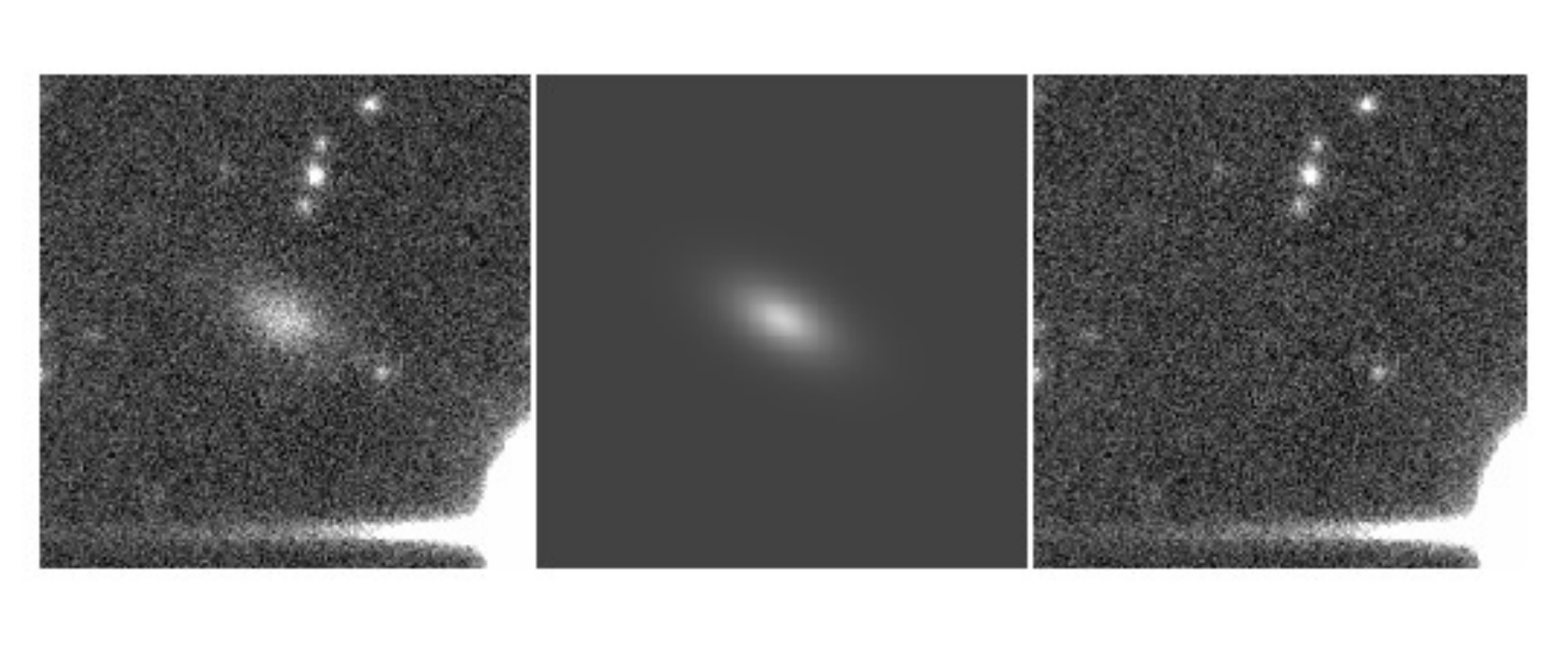}{0.33\textwidth}{(27)}
          }
 \gridline{
          \fig{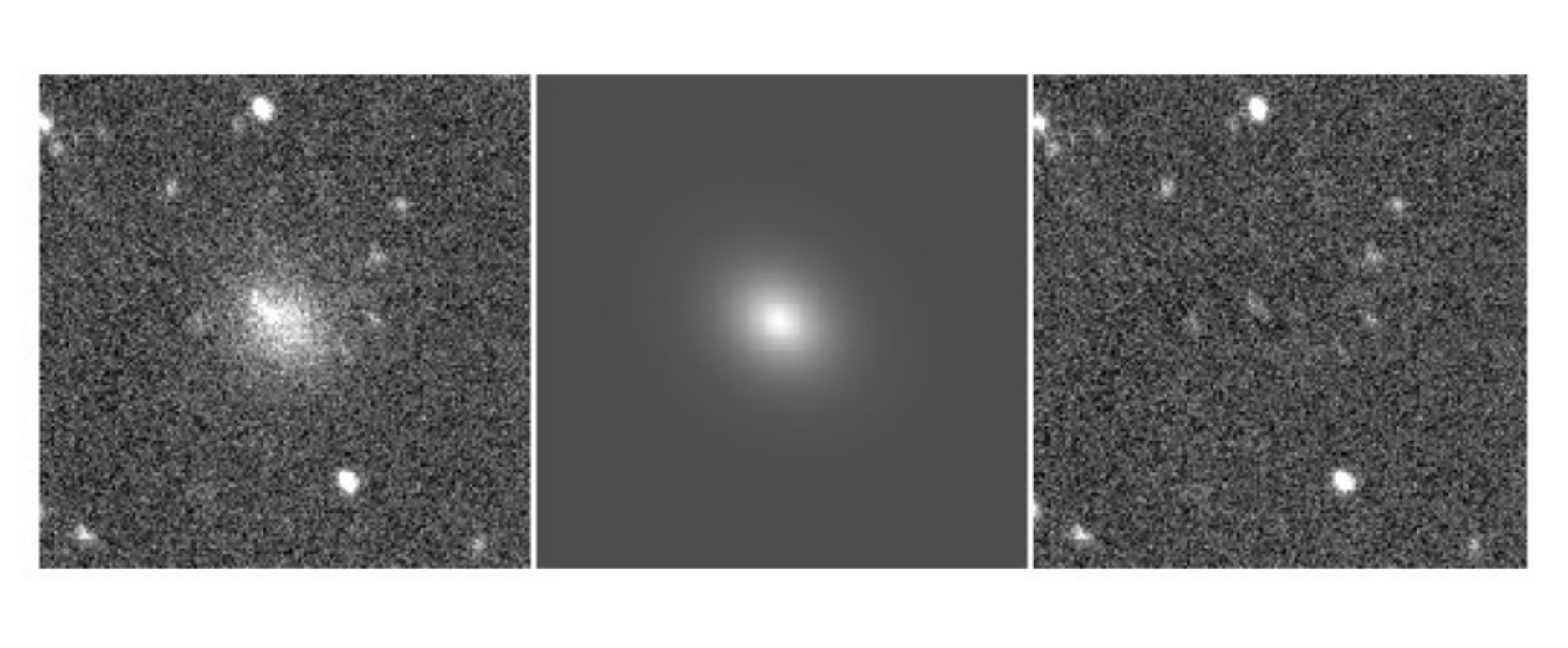}{0.33\textwidth}{(28)}
          \fig{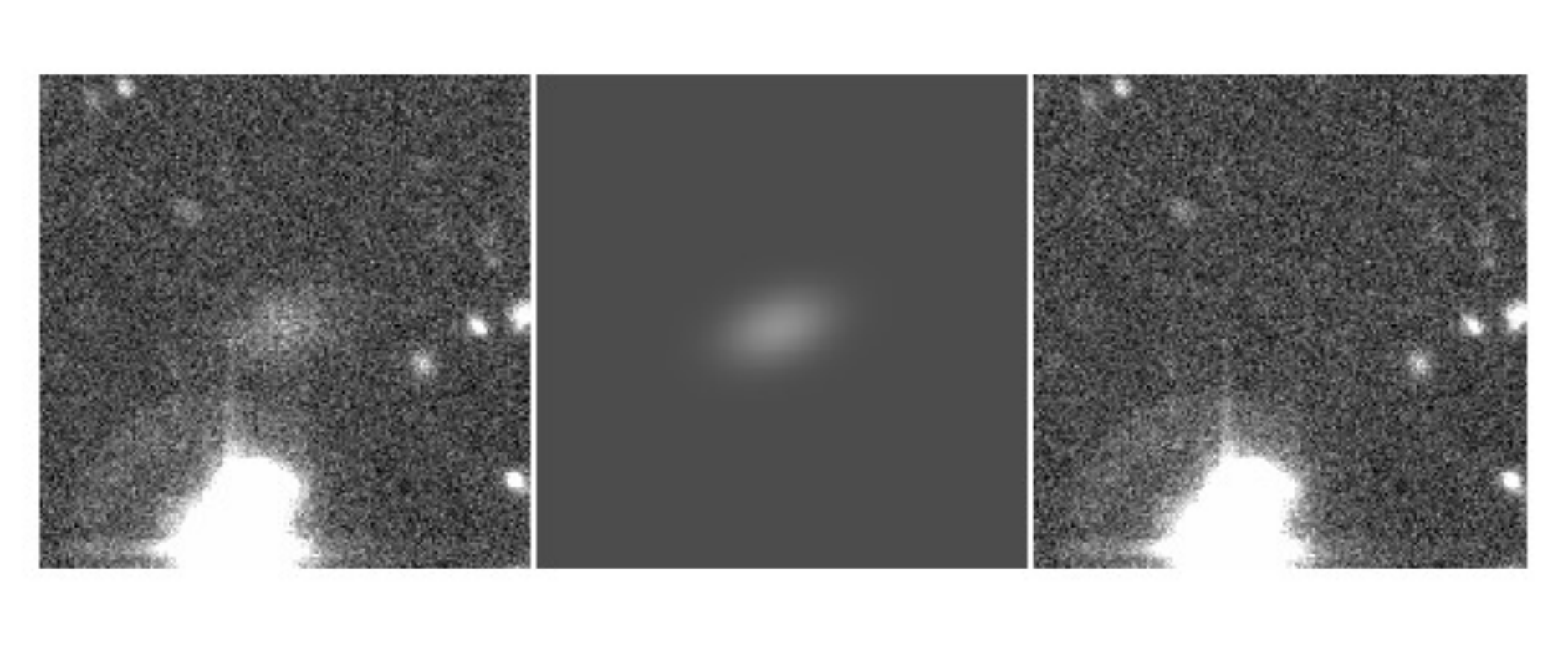}{0.33\textwidth}{(29)}
          \fig{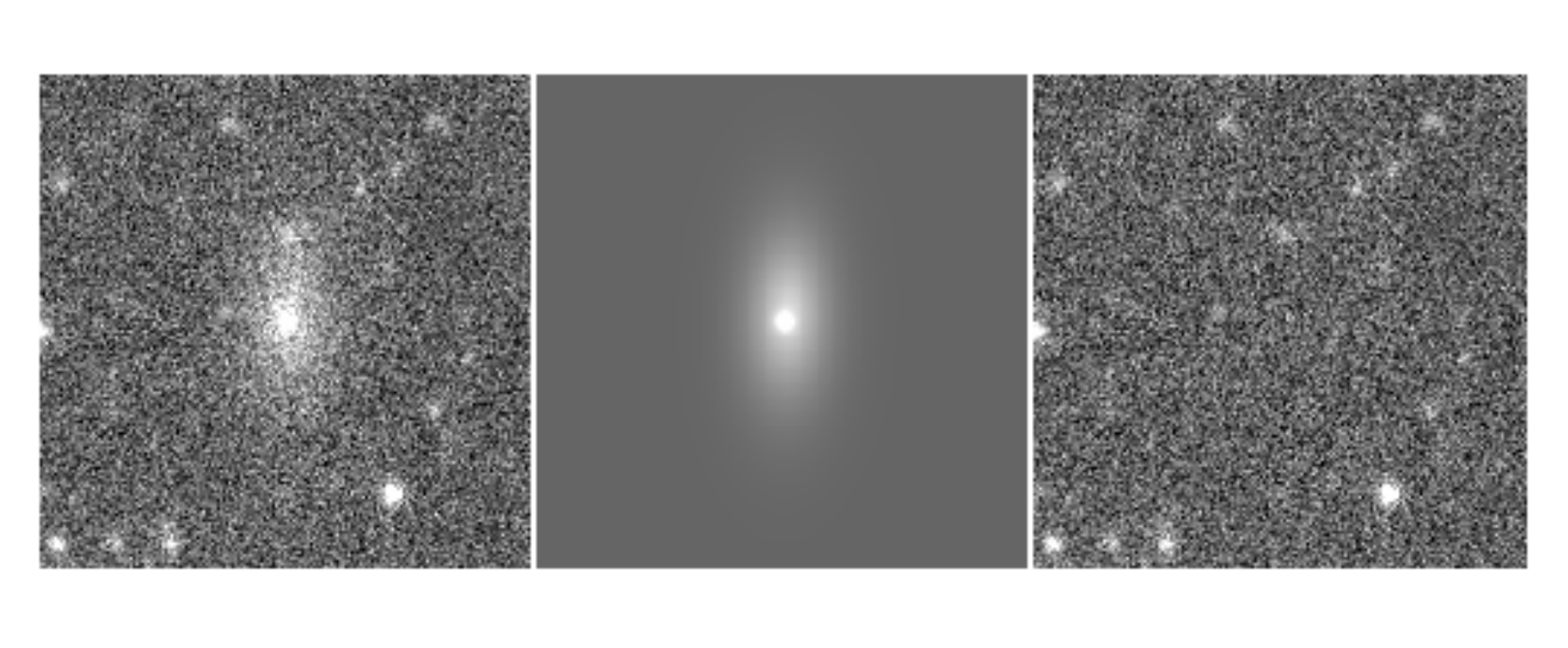}{0.33\textwidth}{(30)}
          }
\gridline{
          \fig{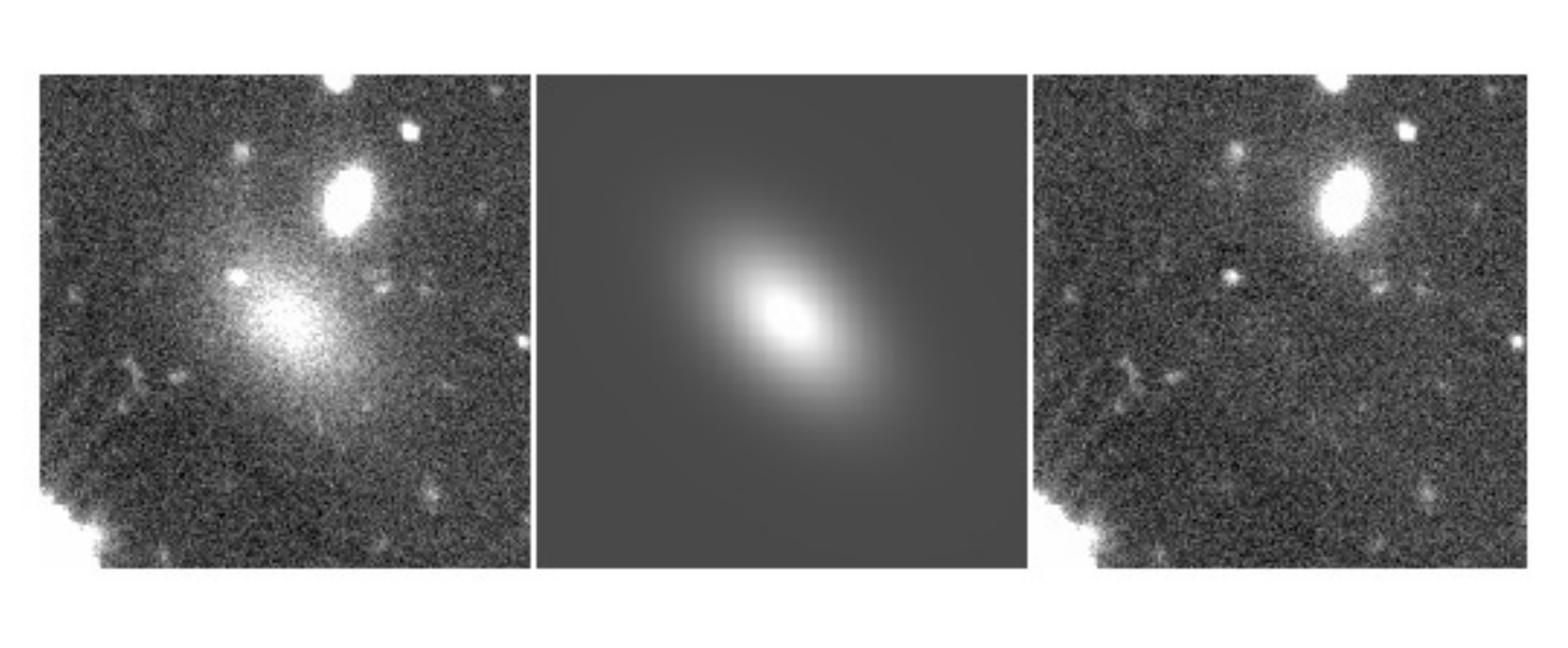}{0.33\textwidth}{(31)}
          \fig{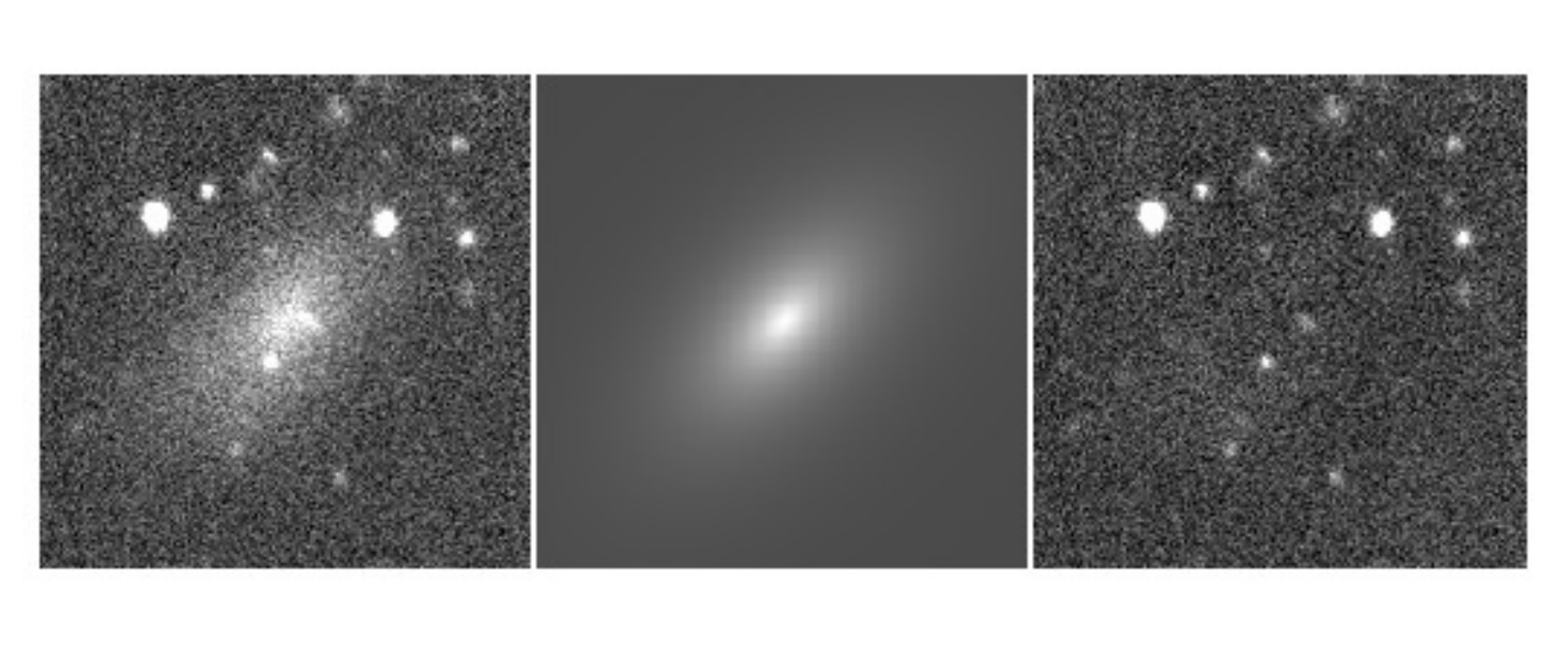}{0.33\textwidth}{(32)}
          \fig{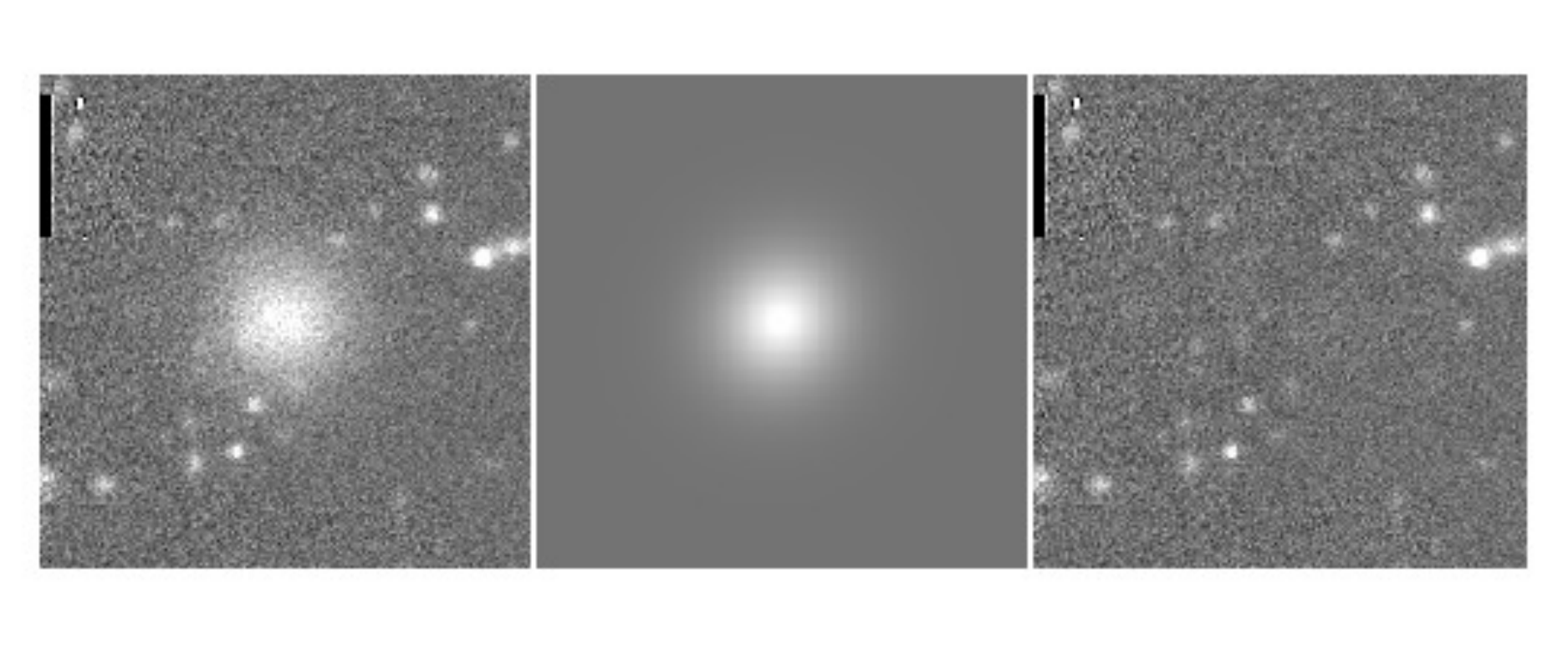}{0.33\textwidth}{(33)}
          }
\gridline{
          \fig{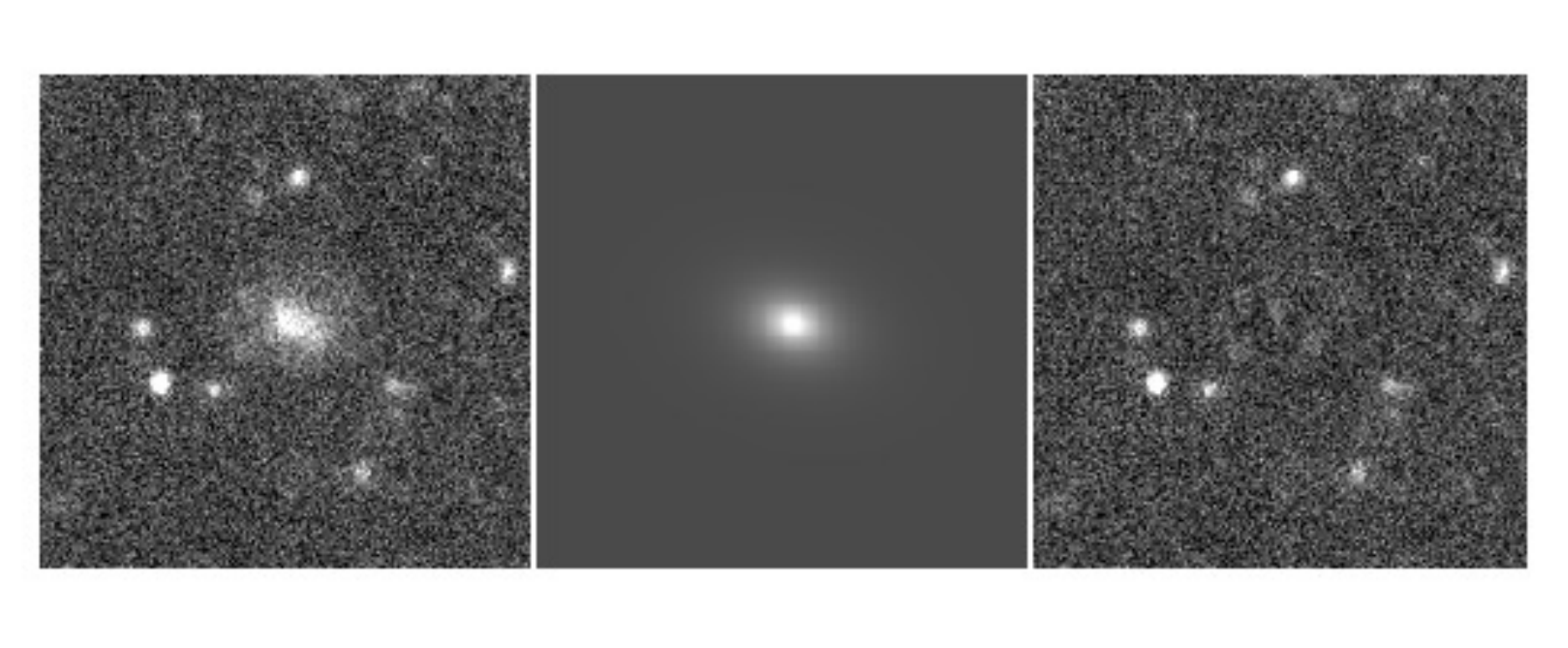}{0.33\textwidth}{(34)}
          \fig{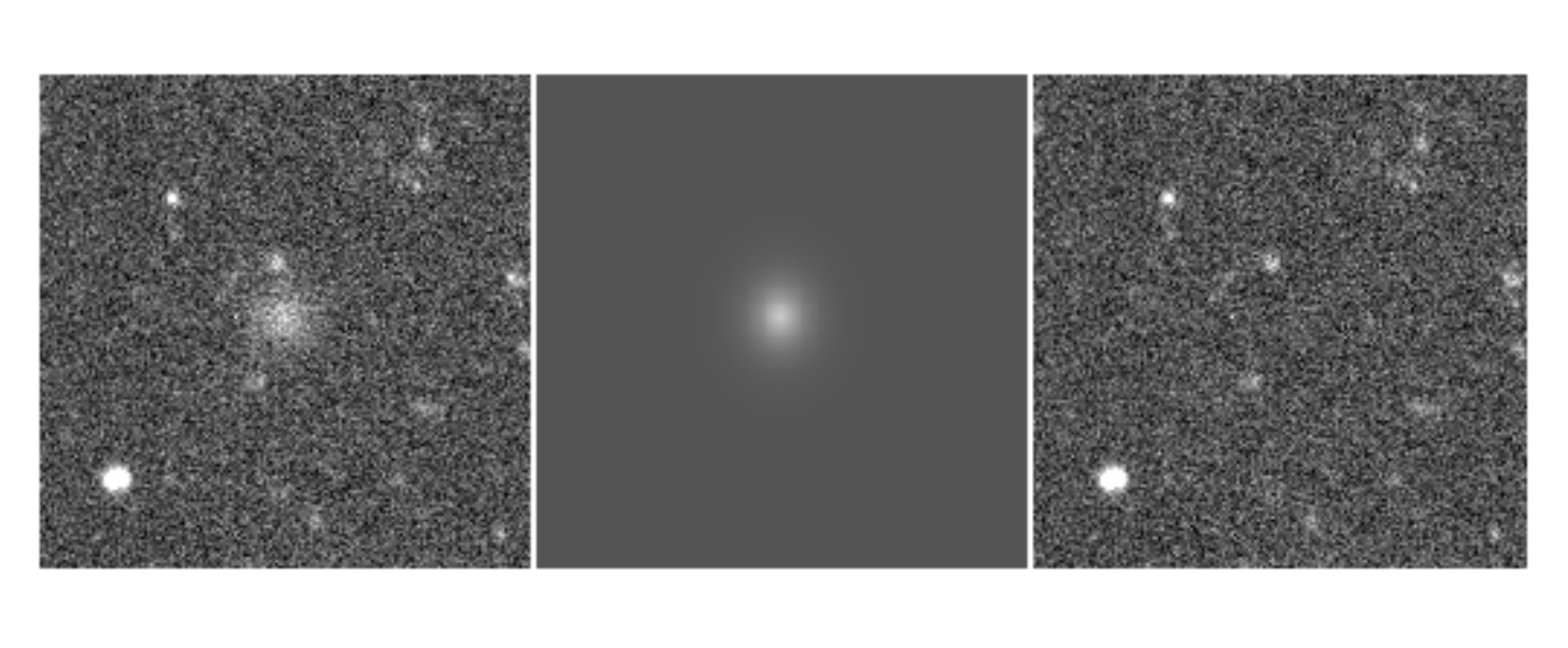}{0.33\textwidth}{(35)}
          \fig{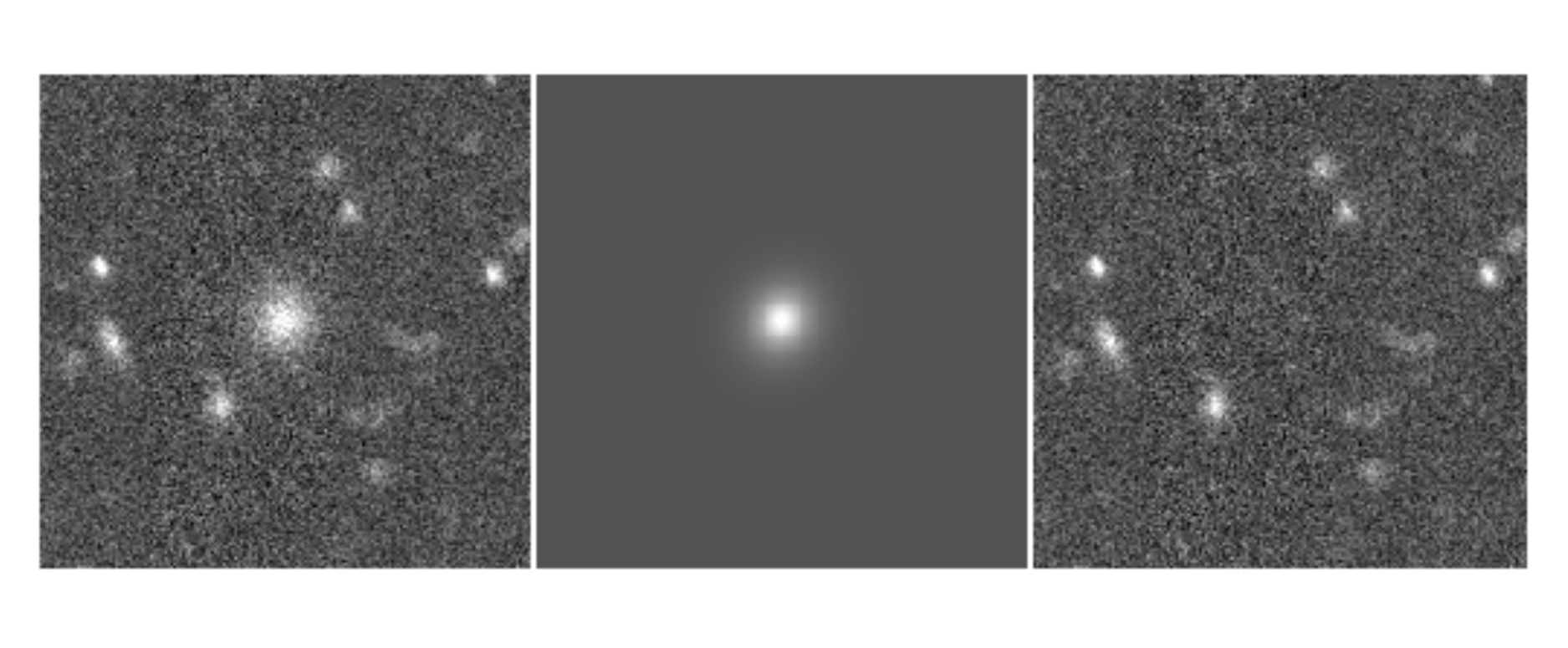}{0.33\textwidth}{(36)}
          }
\gridline{
          \fig{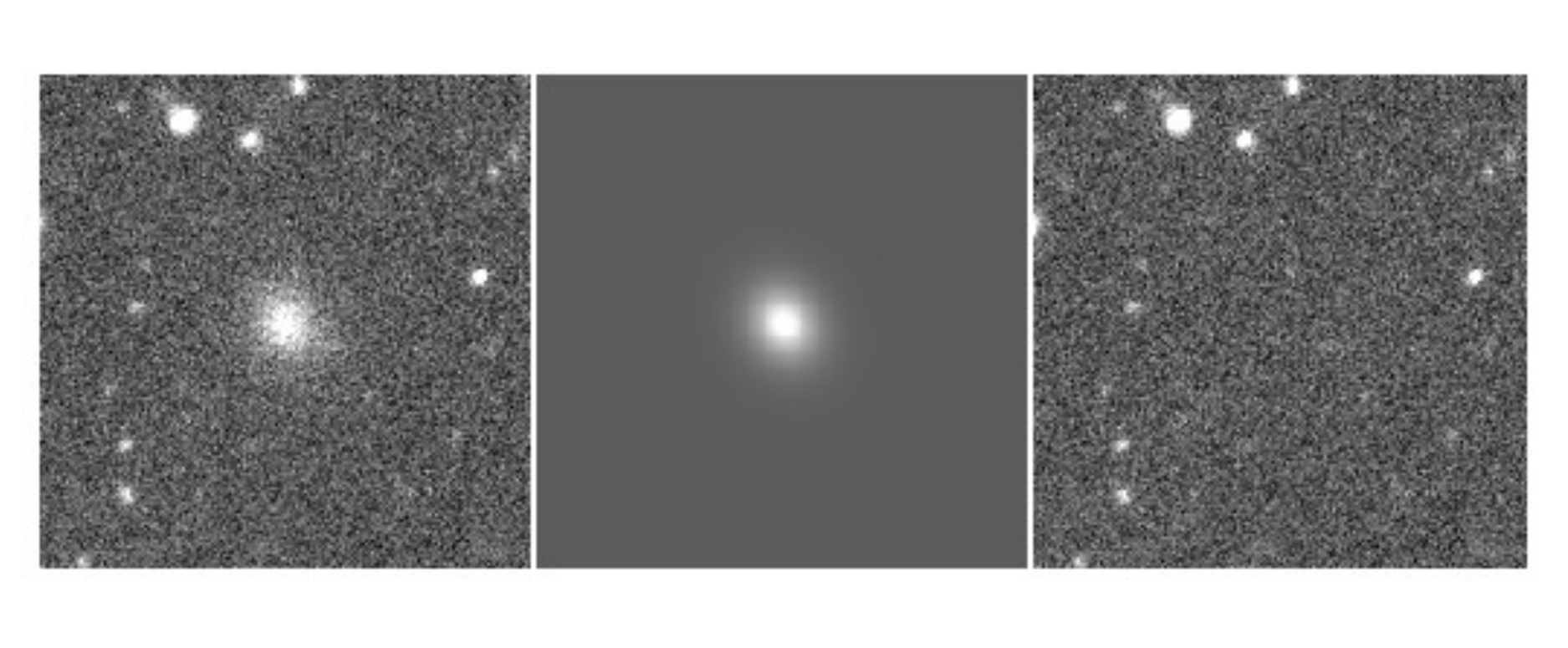}{0.33\textwidth}{(37)}
          \fig{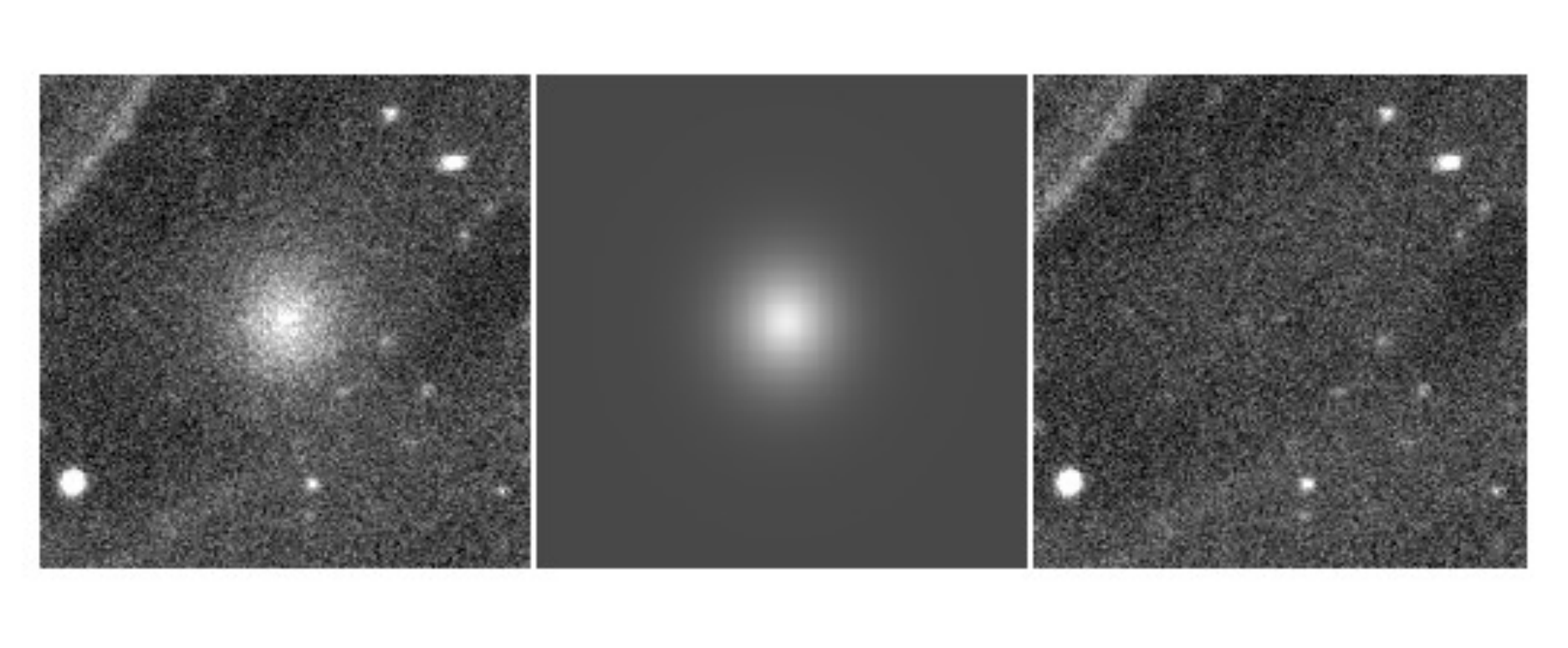}{0.33\textwidth}{(38)}
          }
\caption{The GALFIT models fitted to the new dwarf galaxies. Images are 37.2''x32.7''. North is up, East is Left. Left panel: $g$ band CFHTLS image. Center panel: GALFIT model. Right panel: Residuals.\label{fig:galfit_2}}
\end{figure}

\bibliographystyle{apj}

\bibliography{ref_PB}

\end{document}